\documentclass[letterpaper,reprint,prd,superscriptaddress,nofootinbib,longbibliography]{revtex4-1}

\usepackage[utf8]{inputenc}

\usepackage{feynmf}
\usepackage{feyn}
\usepackage{csquotes}
\usepackage{slashed}

\usepackage{rotating}
\usepackage{adjustbox}
\usepackage{xcolor}
\usepackage{mathtools}
\usepackage{braket}
\usepackage{float}
\usepackage{graphicx}

\usepackage{amsmath,tikz}
\usepackage[makeroom]{cancel}
\usepackage{empheq}
\usepackage[theorems,skins]{tcolorbox}
\usetikzlibrary{matrix,calc,decorations.markings}
\usepackage{amssymb}
\usepackage{amsfonts}

\usepackage{indentfirst}
\usepackage{afterpage}
\usepackage[titletoc]{appendix}
\usepackage[top=0.8in,bottom=0.8in,left=0.8in,right=0.8in]{geometry}
\usepackage{dcolumn}
\usepackage{titlesec}
\usepackage[hidelinks]{hyperref}
\hypersetup{
     colorlinks   = true,
     linkcolor = teal,
     urlcolor  = teal,
     citecolor = blue,
}

\def\headline#1{\hbox to \hsize{\hrulefill\ \lower.3em\hbox{#1}\ \hrulefill}}

\allowdisplaybreaks
\usepackage{bbold}

\newcommand{\MP}{M_{\rm Pl}}

\begin{document}
\title{Equivalence of inflationary models between\\ the metric and Palatini formulation of scalar-tensor theories 
}

\author{Laur J\"arv}
\email{laur.jarv@ut.ee}
\affiliation{Institute of Physics, University of Tartu, W. Ostwaldi 1, 50411 Tartu, Estonia}
\author{Alexandros Karam}
\email{alexandros.karam@kbfi.ee}
\affiliation{National Institute of Chemical Physics and Biophysics, R\"avala 10, 10143 Tallinn, Estonia}
\author{Aleksander Kozak}
\email{aleksander.kozak@uwr.edu.pl}
\affiliation{Institute of Theoretical Physics University of Wroclaw, pl. Maxa Borna 9, 50-206 Wroclaw, Poland}
\author{Angelos Lykkas}
\email{a.lykkas@uoi.gr}
\affiliation{Physics Department, University of Ioannina, GR–45110 Ioannina, Greece}
\author{Antonio Racioppi}
\email{antonio.racioppi@kbfi.ee}
\affiliation{National Institute of Chemical Physics and Biophysics, R\"avala 10, 10143 Tallinn, Estonia}
\author{Margus Saal}
\email{margus.saal@ut.ee}
\affiliation{Institute of Physics, University of Tartu, W. Ostwaldi 1, 50411 Tartu, Estonia}


\begin{abstract}
With a scalar field non-minimally coupled to curvature, the underlying geometry and variational principle of gravity -- metric or Palatini -- becomes important and makes a difference, as the field dynamics and observational predictions generally depend on this choice. In the present paper we describe a classification principle which encompasses both metric and Palatini models of inflation, employing the fact that inflationary observables can be neatly expressed in terms of certain quantities which remain invariant under conformal transformations and scalar field redefinitions. This allows us to elucidate the specific conditions when a model yields equivalent phenomenology in the metric and Palatini formalisms, and also to outline a method how to systematically construct different models in both formulations that produce the same observables.
\end{abstract}

\maketitle

\section{Introduction}
Recent observations of the cosmic microwave background radiation (CMB) indicate that at large scales the Universe is flat and homogeneous. These features can be explained by postulating a quasi-de Sitter expansion during the very early moments of the Universe. Furthermore, this inflationary era is able to generate and preserve the primordial inhomogeneities which became the seeds for the subsequent large-scale structure that we observe. Inflation is usually formulated by supplementing the Einstein-Hilbert action with one or more real scalar fields whose energy density drives the near-exponential expansion. 

Recently, the Planck satellite mission~\cite{Akrami2018} has constrained the available parameter space and essentially excluded many inflationary models. Two of the most popular models, namely Starobinsky~\cite{Starobinsky1980} and non-minimal Higgs inflation~\cite{Bezrukov2008, Bezrukov2009a, Bezrukov2009, Bezrukov2011} still lie in the allowed region. Incidentally, these theories, even though seemingly very different, belong to the same equivalence class which is why they give the same predictions for the observables. They also belong to the class of scalar-tensor theories where the inflaton is generally non-minimally coupled to gravity but minimally coupled to matter (Jordan frame). Of course, one can always perform a rescaling of the metric and a scalar field reparametrization and move to the Einstein frame where the scalar field is minimally coupled to gravity. One can work in either frame, while there is an ongoing debate as to which one is physical~\cite{Quiros2013, Chiba2013a, Kamenshchik:2014waa,Domenech2015a, Herrero-Valea2016, Burns2016, Brooker2016, Bhattacharya2017a, Pandey2017, Bahamonde2017, Karamitsos2017, Ruf2018, Karamitsos2018, Bhattacharya2018, Quiros2018, Falls2018, Chakraborty2019, Quiros2019, Karwan:2018eln, Nandi:2019xlj, Ema:2019fdd, Nashed:2019yto}. 
To circumvent the issue, a frame-invariant approach was developed in~\cite{Jaerv2015,Jaerv2015a, Jarv2015}, then fruitfully applied to slow-roll inflation~\cite{Kuusk2016, Jarv2017, Karam2017} and extended to related theories and formulations~\cite{Kuusk2016a, Karam2018, Kozak:2018vlp, Borowiec:2020lfx, Hohmann2018}. The advantage of this method is that, starting from any scalar-tensor theory, one can define quantities that remain invariant under the conformal Weyl rescaling of the metric and scalar field reparametrization and then express the inflationary observables in terms of these invariants. 

Another issue that arises when one is interested in non-minimally coupled theories is that of the employed variational principle. In the metric formalism, the metric is the only dynamical degree of freedom and the connection is the Levi-Civita. However, in the Palatini or first order formalism~\cite {Palatini1919, Ferraris1982}, the metric and the connection are assumed to be independent
variables and one has to vary the action with respect to both of them. Both approaches lead to the same field equation for an action whose Lagrangian is linear in $R$ and is minimally coupled, but this is no longer true for more general actions. Regarding inflation, the difference in the predictions between the two variational principles has been recently studied in~\cite{Exirifard2008, Bauer2008, Bauer2011, Tamanini2011, Bauer:2010jg, Olmo2011, Rasanen2017, Racioppi2017, Racioppi2018, Jaerv2018, Kozak:2018vlp, Bombacigno2019, Rasanen2019, Rasanen2018,  Almeida2019, Shimada2019, Takahashi2019, Jinno2019, Tenkanen2019, Edery2019, Jinno2020, Aoki2019, Giovannini2019, Tenkanen2019a, Bostan2019a, Bostan2019, Tenkanen2020, Gialamas2020, Racioppi2019, Antoniadis2018, Antoniadis2019, Antoniadis2019a, Tenkanen2020a, Tenkanen2020b, LloydStubbs2020, Antoniadis2020, Borowiec:2020lfx, Ghilencea2020}. 
In most of the previous studies it was shown that the metric and Palatini formulations generally give different results when inflation is concerned (see however~\cite{Racioppi2017, Racioppi2018}). In this article we focus on the cases when the two formalisms can produce similar results and extend the classification scheme of~\cite{Jarv2017} to include Palatini models. 

Future space missions (LITEBIRD \cite{Matsumura2016}, PIXIE \cite{Kogut_2011}, PICO \cite{Sutin:2018onu}) promise to determine the inflationary observables at high precision that will considerably narrow the range of viable models. However, even when the invariant potential can be effectively pinned down, there will remain a degeneracy, as many fundamental actions in different formulations and parametrizations can lead to the same invariant potential and hence to the same values for the observables. The aim of the current paper is to clarify the situation and to outline a method of how to explore and reconstruct such equivalent actions in a systematic way. In the end, some actions in a given equivalence class would be better motivated from the theoretical point of view, while the degeneracy could be also broken by some observations of noninflationary physics.

The paper is organized as follows. In the next section we adopt the 
approach of invariants to study general scalar-tensor theories in both metric and Palatini formalisms. In section~\ref{sec:slow-roll} we focus on inflation and express the slow-roll parameters and inflationary observables in terms of the invariant 
potential and its derivatives. Then, in section~\ref{sec:same_action} we determine under which conditions the metric and Palatini formalisms can generate the same slow-roll parameters when one starts from the same action and study some examples. Conversely, starting from the same invariant potential in section~\ref{sec:different_action} we
explore the reconstruction of the corresponding metric and Palatini actions. We summarize our results and conclude in section ~\ref{sec:conclusions}. Finally, we include an appendix~\ref{appendix} where we illustrate how an additional independent (conformal) transformation of the connection enlarges the general Palatini action, but a suitable choice neutralizes the effect, a point that has not received much attention in the literature so far.

\section{Action and invariant quantities}
\label{sec:action}
Regardless the gravity formulation, the action for general scalar-tensor theory can be written as\footnote{The most general Palatini action contains also additional terms due to the non-metricity of the theory~\cite{Kozak:2018vlp}. However, it is possible to show that the action can always be cast in the form of eq.~\eqref{eq:action}. For the interested reader the concerning details are given in Appendix \ref{appendix}.} \cite{Flanagan:2004bz},
\begin{align}
\mathcal{S}=&\int\!\mathrm{d}^4x\,\sqrt{-g}\left\{\frac{1}{2}\mathcal{A}(\Phi)R-\frac{1}{2}\mathcal{B}(\Phi)\left(\nabla\Phi\right)^2-\mathcal{V}(\Phi)\right\}
\nonumber\\
&  \qquad + \mathcal{S}_m[\mathrm{e}^{2\sigma(\Phi)} g_{\mu\nu},  \chi_{m}] \, , \label{eq:action}
\end{align}
where we used Planck units $\MP=1$ and metric signature $(-,+,+,+)$. The Ricci scalar $R=g^{\mu\nu}R_{\mu\nu}[\Gamma,\partial\Gamma]$ is a function of the metric tensor $g_{\mu\nu}$ and the connection $\Gamma$. The choice of the gravity formulation is reflected on the expression of $\Gamma$ in eq.~\eqref{eq:action} \cite{Bauer2008}:
\begin{eqnarray}
\Gamma^{\lambda}_{\alpha \beta} &=& 
\left\lbrace {}^{\ \lambda}_{\alpha\beta} \right\rbrace
+ \label{eq:conn:J} \\
&&(1-\delta_{j\,\Gamma}) \left[\delta^{\lambda}_{\alpha} \partial_{\beta} \omega(\Phi) +
\delta^{\lambda}_{\beta} \partial_{\alpha} \omega(\Phi) - g_{\alpha \beta} \partial^{\lambda}  \omega(\Phi) \right] \, , \nonumber
\end{eqnarray}
where 
\begin{eqnarray}
\label{omega}
\omega\left(\Phi\right)=\ln\sqrt{\mathcal{A}(\Phi)} \, ,
\end{eqnarray}
$\lbrace {}^{\ \lambda}_{\alpha\beta} \rbrace$ is the Levi-Civita connection, $\delta_{jk}$ is the Kronecker delta and $j=g$ stands for the metric case while  $j = \Gamma$ for the Palatini one.

We refer to the set of $\left\{\mathcal{A},\,\mathcal{B},\,\mathcal{V},\,\sigma\right\}$ as the model functions. By considering a Weyl rescaling of metric (referred later as a change of frame) and scalar field redefinition (referred later as a reparametrization)
\begin{subequations}
	\label{Weyl_transformation_and_field_reparametrization}
	\begin{align}
	\label{conformal_transformation}
	g_{\mu\nu} &= \mathrm{e}^{2\bar{\gamma}(\bar{\Phi})}\bar{g}_{\mu\nu} \,, \\
	\label{field_redefinition}
	\Phi &= \bar{f}(\bar{\Phi}) \,,
	\end{align}
\end{subequations}
the action functional \eqref{eq:action} preserves its structure (up to the boundary term) if the functions $\mathcal{A}$, $\mathcal{B}$, $\mathcal{V}$ and $\sigma$ transform as \cite{Flanagan:2004bz}
\begin{subequations}
	\label{fl_fnide_teisendused}
	\begin{align}
	\label{A_transformation}
	\bar{\mathcal{A}}(\bar{\Phi}) &= \mathrm{e}^{2\bar{\gamma}(\bar{\Phi})}
	{\mathcal A} \left( {\bar f}( {\bar \Phi})\right) \,,\\
	\label{B_transformation}
	{\bar {\mathcal B}}({\bar \Phi}) &= \mathrm{e}^{2{\bar \gamma}({\bar \Phi})} \,
	\left(\bar{f}^\prime\right)^2{\mathcal B} (\bar{f}(\bar{\Phi}))  
	\nonumber \\
	&\quad -  6 \, \delta_{j\,\Gamma} \, \mathrm{e}^{2{\bar \gamma}({\bar \Phi})} \left[
	\left(\bar{\gamma}^{\,\prime}\right)^2 {\mathcal A} \left(\bar{f}(\bar{\Phi})\right) -
	\bar{\gamma}^{\,\prime}\bar{f}^\prime \mathcal{A}^\prime \right] \,, \\
	\label{V_transformation}
	\bar{{\mathcal V}}(\bar{\Phi}) &= \mathrm{e}^{4\bar{\gamma}(\bar{\Phi})} \, {\mathcal V}\left(\bar{f}(\bar{\Phi})\right) \,, \\
	\label{s_transformation}
	\bar{\sigma}(\bar{\Phi}) &= \sigma \left(\bar{f}(\bar{\Phi})\right) + \bar{\gamma}(\bar{\Phi})\, ,
	\end{align}
\end{subequations}
where prime denotes a derivative with respect to the scalar field.
The Jordan frame is defined by the condition $\sigma(\Phi) = 0$. For what follows we omit the matter part of the action and take $\mathcal{S}_{m} = 0$, since our interest is now on the scalar non-minimally coupled to gravity which will be identified with the inflaton field.

By a straightforward calculation it is possible to make sure, that 
in every spacetime point the numerical value of the quantities~\cite{Jaerv2015}
\begin{align}
	\label{I.1}
    \mathcal{I}_{\mathrm{m}}(\Phi) &= \frac{\mathrm{e}^{2\sigma(\Phi)}}{\mathcal{A}(\Phi)},\\
	\label{I.2}    
    \mathcal{I}_{\mathcal{V}}(\Phi) &= \frac{\mathcal{V}(\Phi)}{\left(\mathcal{A}(\Phi)\right)^2},\\
	\label{I.3} 
	\mathcal{I}_{\Phi}(\Phi) &= \int\!\mathrm{d}\Phi\,\sqrt{\frac{\mathcal{B}(\Phi)}{\mathcal{A}(\Phi)} 
		+  \frac{3}{2} \, \delta_{j\,\Gamma}  \, \left(\frac{\mathcal{A}'(\Phi)}{\mathcal{A}(\Phi)}\right)^2}
\end{align} 
remain invariant, i.e. 
$\bar{\mathcal{I}}_{i}(\bar{\Phi}) = \mathcal{I}_{i}(\Phi)$. In a similar vein, we may introduce an invariant metric $\hat{g}_{\mu\nu}=\mathcal{A}\,g_{\mu\nu}$, which is unaffected by the conformal transformation \eqref{conformal_transformation}. 
One can see that the invariant field $\mathcal{I}_{\Phi}$ has a different dependence on the model functions when one considers the metric (we use the notation $\mathcal{I}_{\Phi}^{g}$) 
or Palatini formalism (denoted as $\mathcal{I}_{\Phi}^{\Gamma}$). 
Still, in both formalisms we may take the quantity $\mathcal{I}_\Phi$ as an invariant description of the scalar degree of freedom in the theory \cite{Jaerv2015, Kozak:2018vlp}. Negative values for the expression under the square root in eq.~\eqref{I.3} suggest that the scalar field is a ghost, while identically constant $\mathcal{I}_\Phi$ indicates that the scalar is not dynamical. In the metric formulation this occurs only when $\mathcal{B} (\Phi) = -\frac{3}{2} \frac{\left(\mathcal{A}'(\Phi)\right)^2}{\mathcal{A}(\Phi)}$, while in the Palatini for $\mathcal{B} (\Phi) =0$. In both cases the theory is equivalent to general relativity plus a cosmological constant. A multiscalar generalization of the integrand in eq.~\eqref{I.3} plays the role of the invariant volume form on the space of scalar fields, hence here $\mathcal{I}_\Phi$ has a natural interpretation as an invariant ``distance'' in the $1$-dimensional space of the scalar field \cite{Kuusk2016a, Kuusk2016}.

By inverting eq.~\eqref{I.3} we may switch to use $\mathcal{I}_{\Phi}$ as the basic variable instead of $\Phi$, and employing the invariant metric $\hat{g}_{\mu\nu}$ we can rewrite the action \eqref{eq:action} in terms of invariant quantities only \cite{Jaerv2015},
\begin{align}
    \mathcal{\hat{S}}=&\int\!\mathrm{d}^4x\,\sqrt{-\hat{g}}\left\{\frac{1}{2}\hat{g}^{\mu\nu}R_{\mu\nu}[\Gamma, \hat{g}_{\mu\nu}]-\frac{1}{2}(\hat{\nabla}\mathcal{I}_{\Phi})^2 -\mathcal{I}_{\mathcal{V}} \right\}\nonumber\\
    &\qquad+\mathcal{S}_m[\mathcal{I}_m\,\hat{g}_{\mu\nu},\psi] \,.
    \label{Einstein.frame.action}
\end{align}
An arbitrary scalar-tensor theory with four free functions \eqref{eq:action} can therefore be cast by the two transformations of frame change and reparametrization \eqref{Weyl_transformation_and_field_reparametrization} into the action \eqref{Einstein.frame.action} endowed by two functions that carry invariant meaning. The quantity $\mathcal{I}_{\mathrm{m}}(\mathcal{I}_{\Phi})$ characterizes the coupling of gravity to matter fields. For constant $\mathcal{I}_{\mathrm{m}}$ the theory is equivalent to general relativity with a minimally coupled scalar field, otherwise the scalar field participates in mediating the gravitational interaction and the effective gravitational ``constant'' starts to vary according to the scalar field value. The quantity $\mathcal{I}_{\mathcal{V}}(\mathcal{I}_{\Phi})$ is the invariant scalar potential. In the case of inflation where the matter fields can be neglected, the physics of the model is encoded by the invariant potential alone \cite{Jarv2017}. The form of the invariant action \eqref{Einstein.frame.action} coincides with the usual Einstein frame action, a circumstance which will help us to write down the inflationary parameters in terms of the invariants in the next section.

\section{Slow-roll parameters and computational algorithm}
\label{sec:slow-roll}

The action functional \eqref{Einstein.frame.action} can be identified as the Einstein frame regarding the $\hat{g}_{\mu\nu}$ metric. Then the equations of motion coincide in both formulations of gravity, although in the Palatini formalism the Levi-Civita connection is derived on-shell from its constraint equation $\delta_{(\Gamma)}\hat{\mathcal{S}}=0$. The invariant quantity $\mathcal{I}_\Phi$ assumes the role of the inflaton field driving inflation, governed by its potential $\mathcal{I_V}(\mathcal{I}_\Phi)$. Assuming then the usual slow-roll conditions, we can rewrite the potential slow-roll parameters (PSRPs) as~\cite{Kuusk2016, Jarv2017, Karam2017}
\begin{align}
	\label{epsilon}
	\epsilon &= \frac{1}{2}\left(\frac{\mathrm{d} \ln \mathcal{I}_{\mathcal{V}}}{\mathrm{d} \mathcal{I}_{\Phi}}\right)^2\,,\\
	\label{eta}
	\eta &= \frac{1}{\mathcal{I}_{\mathcal{V}}}\frac{\mathrm{d}^2 \mathcal{I}_{\mathcal{V}}}{\mathrm{d} \mathcal{I}_{\Phi}^2}\,.
\end{align}
At this point, we assumed that the integral in eq.~\eqref{I.3} is solvable and the relation of $\mathcal{I}_\Phi(\Phi)$ invertible\footnote{The problem is still solvable also when $\mathcal{I}_\Phi(\Phi)$ is not invertible. In that case $\Phi$ is used as a \emph{new} variable and the chain rule is applied in the computation of the derivatives.}, so that we can obtain a relation of $\Phi(\mathcal{I}_\Phi)$. Then, after a direct substitution into $\mathcal{I_V}$ we express the PSRPs in terms of $\mathcal{I_V}(\mathcal{I}_\Phi)$. 

The tensor-to-scalar ratio $r$, the scalar spectral index $n_s$ and the amplitude of the scalar power spectrum $A_s$ are some of the inflationary observable quantities posing strict constraints on the parameter space of the inflationary models. These are usually computed in the slow-roll approximation and, up to first order in PSRPs, they read as follows~\cite{Jarv2017, Karam2017}:
\begin{align}
	\label{r}
	r &= 
	8\left(\frac{\mathrm{d} \ln \mathcal{I}_{\mathcal{V}}}{\mathrm{d} \mathcal{I}_{\Phi}}\right)^2 \,,
	\\
	\label{ns}
	n_s &= 
	1 -3 \left(\frac{\mathrm{d} \ln \mathcal{I}_{\mathcal{V}}}{\mathrm{d} \mathcal{I}_{\Phi}}\right)^2+2\,\frac{1}{\mathcal{I}_{\mathcal{V}}}\frac{\mathrm{d}^2 \mathcal{I}_{\mathcal{V}}}{\mathrm{d} \mathcal{I}_{\Phi}^2}  \,,
	\\
	\label{scalar.amplitude}
	A_s &= \frac{\mathcal{I}_{\mathcal{V}}}{12\pi^2}\left(\frac{\mathrm{d} \ln \mathcal{I}_{\mathcal{V}}}{\mathrm{d} \mathcal{I}_{\Phi}}\right)^{-2} \,.
\end{align}
Note that all of the above observables are calculated at horizon exit, $\mathcal{I}_\Phi=\mathcal{I}_\Phi^*$. The number of e-foldings, characterizing the duration of inflation, is given by
\begin{equation}
	\label{number.of.efolds}
	N= \int\limits_{\mathcal{I}_{\Phi}^\text{end}}^{\mathcal{I}_{\Phi}^*}
	\mathcal{I}_{\mathcal{V}}(\mathcal{I}_{\Phi})\left(\frac{\mathrm{d}\mathcal{I}_{\mathcal{V}}(\mathcal{I}_{\Phi})}{\mathrm{d}\mathcal{I}_{\Phi}}\right)^{-1}\mathrm{d} \mathcal{I}_{\Phi}\,,
\end{equation}
where $\mathcal{I}_{\Phi}^\text{end}$ and $\mathcal{I}_{\Phi}^*$ are the field values at the end and start of inflation, respectively. 

The invariant formalism can be applied in a straightforward way to any model that can be recast in the form of eq.~\eqref{Einstein.frame.action}, by first identifying the model functions $\mathcal{A}(\Phi)$, $\mathcal{B}(\Phi)$, $\mathcal{V}(\Phi)$ and $\sigma(\Phi)$. This is under the implicit assumption that the model under consideration includes only one dynamical scalar field $\Phi$. As we explained previously, we may use \eqref{I.3} to compute the invariant quantity $\mathcal{I}_\Phi(\Phi)$ and invert that relation to obtain $\Phi(\mathcal{I}_\Phi)$. By using $\Phi(\mathcal{I}_\Phi)$ we can calculate the invariant potential $\mathcal{I_V}(\mathcal{I}_\Phi)$ and then solve $\epsilon(\mathcal{I}_{\Phi}^\text{end})=1$ to obtain the field value at the end of inflation. The field value $\mathcal{I}_\Phi^*$ is obtained by integrating \eqref{number.of.efolds} and assuming that the number of e-folds lies somewhere in the allowed region of $N\simeq(50-60)$ e-folds. Finally, the inflationary observables are readily obtained from eqs.~\eqref{r}-\eqref{number.of.efolds} using the field value $\mathcal{I}_\Phi^*$.

In the following sections, we apply this procedure in the study of the inflationary predictions for scalar-tensor theories in the metric and Palatini formulations.

\section{When do identical metric and Palatini actions yield (almost) the same observables?}
\label{sec:same_action}

Comparing eqs.~\eqref{I.2} and~\eqref{I.3} we can see that the difference between the metric and Palatini formulation arise from a different definition of the invariant field value. Therefore, given the action in eq.~\eqref{eq:action} (i.e. a set of functions $\mathcal{A}$, $\mathcal{B}$ and $\mathcal{V}$), the metric and Palatini formulations usually generate different invariant actions and therefore different predictions. However, it might happen that the two formulations produce the same slow-roll parameters when the invariant potential and the invariant field value possess certain properties.
The slow-roll parameters are independent of the overall normalization of the invariant potential, therefore it is enough to assume that invariant potential in the two formulations, as functions of the corresponding invariant field values, are proportional to each other:
\begin{equation}
 \mathcal{I}^{\mathrm{g}}_{\mathcal V}  \propto \mathcal{I}^{\mathrm{\Gamma}}_{\mathcal V} \, .
 \label{eq:IVmproptoIVP}
\end{equation}
Unfortunately, we cannot provide a general criterion that is more explicit than eq.~\eqref{eq:IVmproptoIVP}, because eq.~\eqref{I.3} contains an integral over $\Phi$ and the corresponding solving technique is strongly dependent on the actual definition of $\mathcal{A}$ and $\mathcal{B}$. On the other hand, we can provide a couple of explicit examples: one relatively simple (A) and one more complicated (B).

\subsection{Example: Power law invariant potential}
Given eqs.~\eqref{I.2} and \eqref{I.3}, the simplest way to satisfy eq.~\eqref{eq:IVmproptoIVP} is by requiring
\begin{align}
 \mathcal{I}_{\mathcal{V}} &\propto  \mathcal{I}_{{\Phi}}^n \, , \label{eq:IVmproptoIphin}\\
 \mathcal{I}^{\mathrm{g}}_{\Phi} &\propto  \mathcal{I}^{\mathrm{\Gamma}}_{\Phi} \label{eq:IphimproptoIphiP}\, ,
\end{align} 
where $n$ is some nonzero power. The class of models \eqref{eq:IVmproptoIphin} is well-known (e.g. \cite{Akrami2018}) and goes under the name of monomial inflation.
Using eqs.~\eqref{epsilon},~\eqref{eta} and~\eqref{eq:IVmproptoIphin} we see that the corresponding slow-roll parameters are
\begin{align}
	\epsilon &=  \frac{n^2}{2 \mathcal{I}_{\Phi}^2}\, ,\\
	\eta &=  \frac{n (n-1)}{\mathcal{I}_{\Phi}^2}\, .
\end{align}
We can appreciate that the case $n=2$ is even more special, since it accidentally implies also that $\epsilon = \eta$. 
With a couple of straightforward computations, we can easily verify that the tensor-to-scalar ratio and the scalar spectral index are:
\begin{equation}
r = \frac{16n}{n + 4N}.
\end{equation}
\begin{equation}
n_s = 1 - \frac{2(n + 2)}{n + 4N},
\end{equation}

Furthermore, the combination of eqs.~\eqref{I.3} and \eqref{eq:IphimproptoIphiP} implies
\begin{equation}
 \frac{\mathcal{B}(\Phi)}{\mathcal{A}(\Phi)} \propto \left(\frac{\mathcal{A}'(\Phi)}{\mathcal{A}(\Phi)}\right)^2 \, ,
\end{equation}
therefore
\begin{equation}
 \mathcal{I}^{\mathrm{\Gamma,g}}_{\Phi}(\Phi) \propto \int d \Phi \left| \frac{\mathcal{A}'(\Phi)}{\mathcal{A}(\Phi)} \right| = \ln \frac{\mathcal{A}(\Phi)}{\mathcal{A}_0} \, ,
\end{equation}
where $\mathcal{A}_0$ is a constant of integration that does not carry any physical meaning and can be used to conveniently set the zero-value of the invariant field according to the problem at hand. Imposing eq.~\eqref{eq:IVmproptoIphin} we obtain
\begin{equation}
\left( \ln \frac{\mathcal{A}(\Phi)}{\mathcal{A}_0} \right)^n \propto \frac{\mathcal{V}(\Phi)}{\mathcal{A}(\Phi)^2} \, .
\end{equation}
Therefore, the metric and Palatini formulations produce the same slow-roll parameters when
\begin{align}
\label{constraint_1}
\mathcal{A}(\Phi) \mathcal{B}(\Phi) &\propto  \left( \mathcal{A'}(\Phi) \right)^2 \,, \\
\mathcal{V}(\Phi) &\propto  \mathcal{A}(\Phi)^2 \left( \ln \frac{\mathcal{A}(\Phi)}{\mathcal{A}_0} \right)^n \, .
\label{constraint_2} 
\end{align}

From the two equations, we can immediately see that the following class of non-minimal Coleman-Weinberg models where\footnote{Without loss of generality we assume $\Phi>0$ in Section \ref{sec:same_action} and \ref{sec:different_action}.}
\begin{align}
\label{eq:Coleman-Weinberg A}
\mathcal{A} (\Phi) &= \xi \Phi^2 \,, \\
\mathcal{B} (\Phi) &= 1 \,, \\
\label{eq:Coleman-Weinberg V}
\mathcal{V} (\Phi) &= \beta \left( \ln \frac{\Phi}{\Phi_0} \right)^n \Phi^4\,,
\end{align} 
satisfies the conditions \eqref{constraint_1} and \eqref{constraint_2}, and therefore generates slow-roll parameters that cannot discriminate between metric and Palatini gravity. These results are in agreement with the findings of \cite{Racioppi2017} and the strong coupling limit of \cite{Racioppi2018}.  

Moreover, eqs.~\eqref{constraint_1} and \eqref{constraint_2} can also be used to back-engineer models. For instance, choosing $n=1$ and a natural inflation potential 
\begin{equation}
\label{eq:exAb_V}
\mathcal{V}(\Phi) = M^4 \left( 1 - \cos\left(\frac{\Phi}{\Phi_0}\right) \right) \, ,
\end{equation}
the addition of the following non-trivial non-minimal coupling to gravity and non-canonical kinetic function
\begin{align}
\label{eq:exAb_A}
\mathcal{A}(\Phi) &= \sqrt{\frac{z}{W\left(z\right)}}\,,\\
\label{eq:exAb_B}
\mathcal{B}(\Phi) &= \frac{\sin ^2\left(\frac{\Phi}{\Phi_0} \right) W\left(z\right)^{3/2}}{4 z^{3/2}   \left(W \left(z\right)+1\right)^2}\,,
\end{align}
where $W(z)$ is the Lambert $W$-function and \mbox{$z=1-\cos \left(\frac{\Phi}{\Phi_0} \right)$}, would generate
\begin{align}
 \mathcal{I}_{\mathcal{V}} &\propto \mathcal{I}_{\Phi} \, , \\
	\epsilon &=  \frac{1}{2 \mathcal{I}_{\Phi}^2}\, ,\\
	\eta &=  0\, ,
\end{align}
regardless of the adopted gravity formulation. Therefore, for $n=1$, the scalar spectral index and the tensor-to-scalar ratio have the following values:
\begin{equation}
\begin{split}
& n_s = 0.9701, \quad r = 0.0796, \quad \text{for } N = 50, \\
& n_s = 0.9751, \quad r = 0.0664, \quad \text{for } N = 60,
\end{split}
\end{equation}
 which are out of the 2$\sigma$ Planck boundaries \cite{Akrami2018}, but still allowed at 3$\sigma$.
Therefore, if the universe happened to be described by a non-minimal scalar field with model functions \eqref{eq:exAb_V}, \eqref{eq:exAb_A}, \eqref{eq:exAb_B} in action \eqref{eq:action}, the slow-roll parameters would not be able to distinguish whether the underlying theory is metric or Palatini in character.

\subsection{Example: Logarithmic invariant potential}
A more complicated way to satisfy eq.~\eqref{eq:IVmproptoIVP} is the following choice:
\begin{align}
 \mathcal{I}_{\mathcal{V}} &\propto  \big( \ln (\mathcal{I}_{{\Phi}}^m) \big)^n \, , \label{eq:IVmproptolnIphin}\\
 \mathcal{I}^{\mathrm{g}}_{\Phi} &\propto  (\mathcal{I}^{\mathrm{\Gamma}}_{\Phi})^l \label{eq:IphimproptoIphim}\, ,
\end{align} 
where $\Phi$ is a subscript while $l, m, n$ are some powers. 
The class of models \eqref{eq:IVmproptolnIphin} can be interpreted as a cosmological constant subject to quantum corrections. Being somehow new (only the case $n=1$ is well-known (e.g. \cite{Akrami2018})), such a class deserves a deeper investigation than the previous example.
Despite the nonlinear relation between the invariant fields in the two formalisms in eq.~\eqref{eq:IphimproptoIphim}, the expressions of the slow-roll parameters coming from eq.~\eqref{eq:IVmproptolnIphin} are
\begin{align}
\epsilon &= \frac{ n^2 }{2 \mathcal{I}_{\Phi}^2 \, \big( \ln \mathcal{I}_{\Phi} \big)^2}\, ,\\
\eta &=  \frac{n  \Big(n-1-\ln \mathcal{I}_{\Phi} \Big) }{\mathcal{I}_{\Phi}^2 \, \big( \ln \mathcal{I}_{\Phi} \big)^2 } \, ,
\end{align}
where the power $m$ canceled out because of properties of the logarithm. Inflation ends when
\begin{equation}
\mathcal{I}^\text{end}_\Phi =\frac{n}{\sqrt{2}}W\left(\frac{n}{\sqrt{2}}\right)^{-1},
\end{equation}
while the number of e-folds turns out to be:
\begin{equation}
\begin{split}
N = \frac{1}{n}\Big[-\frac{\mathcal{I}^2_\Phi}{4} + \frac{\mathcal{I}^2_\Phi}{2}\ln\mathcal{I}_\Phi\Big]^{\mathcal{I}^*_\Phi}_{\mathcal{I}^\text{end}_\Phi}.
\end{split}
\end{equation}
Therefore the value of the invariant field at the horizon crossing is:
\begin{equation}
\mathcal{I}^*_\Phi = \sqrt{4N'W\left(\frac{4N'}{e}\right)^{-1}}, \label{eq:IN:ln}
\end{equation}
where we have defined: 
\begin{equation}
\begin{split}
N' & = n N - \frac{n^2}{8}W\left(\frac{n}{\sqrt{2}}\right)^{-2} \\
& + \frac{n^2}{4}W\left(\frac{n}{\sqrt{2}}\right)^{-2}\ln\left(\frac{n}{\sqrt{2}}W\left(\frac{n}{\sqrt{2}}\right)^{-1}\right).
\end{split}
\end{equation}
Using \eqref{eq:IN:ln}, the scalar spectral index and the tensor-to-scalar ratio can be expressed by:
\begin{equation}
n_s = 1 - \frac{n W\left(\frac{4N'}{e}\right)\left(1 + n + \ln\left(4N' W\left(\frac{4N'}{e}\right)^{-1}\right)\right)}{N'\left(\ln\left(4N' W\left(\frac{4N'}{e}\right)^{-1}\right)\right)^2},
\end{equation}
\begin{equation}
r = \frac{n^2W\left(\frac{4N'}{e}\right) }{2N'\left(\ln\left(4N' W\left(\frac{4N'}{e}\right)^{-1}\right)\right)^2}.
\end{equation}

An example to illustrate this possibility can be realized by the model functions
\begin{align}
\label{eq: ln model A}
\mathcal{A}(\Phi) &= \exp \left(\tfrac{1}{2\sqrt{6}} \, \mathrm{acosh}(2 \Phi) - \tfrac{1}{\sqrt{6}} \Phi \sqrt{4 \Phi^2 -1} \right) \,,\\
\label{eq: ln model B}
\mathcal{B}(\Phi) &= \exp \left(\tfrac{1}{2\sqrt{6}} \, \mathrm{acosh}(2 \Phi) - \tfrac{1}{\sqrt{6}} \Phi \sqrt{4 \Phi^2 -1} \right) \,, \\
\label{eq: ln model V}
\mathcal{V}(\Phi) &= \mathcal{A}(\Phi)^2 \left( \ln \Phi^2 \right)^2 \,.
\end{align}
In this case, by integrating eq.~\eqref{I.3} one obtains the invariant field in the two formalisms as
\begin{align}
\mathcal{I}_\Phi^\Gamma &= \Phi \,, \\
\mathcal{I}_\Phi^g &= \Phi^2 \,.
\end{align} 
While the invariant potentials differ by a constant factor,
\begin{align}
\label{eq: ln model I_V_Gamma}
\mathcal{I}_\mathcal{V}^\Gamma &= 4 \Big( \ln \mathcal{I}_\Phi^\Gamma \Big)^2 \,, \\
\label{eq: ln model I_V_g}
\mathcal{I}_\mathcal{V}^g &= \Big( \ln \mathcal{I}_\Phi^g \Big)^2 \,, 
\end{align}
as expected, the slow-roll parameters coincide:
\begin{align}
\label{eq: ln model epsilon}
\epsilon &=  \frac{4 }{2 \mathcal{I}_{\Phi}^2 \, \big( \ln \mathcal{I}_{\Phi} \big)^2}\, ,\\
\label{eq: ln model eta}
\eta &=  \frac{2  \Big(2-\ln \mathcal{I}_{\Phi} \Big) }{\mathcal{I}_{\Phi}^2 \, \big( \ln \mathcal{I}_{\Phi} \big)^2 }\, 
\end{align}
and therefore yield the same $n_s$ and $r$ (as functions of $\mathcal{I}_\Phi$) in both metric and Palatini formulations (cf. eqs. \eqref{eq:IVmproptolnIphin}, \eqref{eq: ln model epsilon} and \eqref{eq: ln model eta} for $n=2$). In this case the scalar spectral index and the tensor-to-scalar ratio take the following values:
\begin{equation}
\begin{split}
& n_s = 0.9699, \quad r = 0.0556, \quad \text{for } N = 50, \\
& n_s = 0.9752, \quad r = 0.0450, \quad \text{for } N = 60,
\end{split}
\end{equation}
 which are within the 2$\sigma$ Planck boundaries \cite{Akrami2018}.

The model \eqref{eq: ln model A}-\eqref{eq: ln model V} looks rather contrived, but it employs a parametrization where the calculational logic is easy to see. However, hidden somewhere in the infinite possibilities of reparametrizations there might exist a physically better motivated form of the same model, but where the calculations become harder to deal with. It is not easy to guess what a nicer parametrization could be, but as an extra illustration let us just perform a simple scalar field redefinition
\begin{equation}\label{mapto}
\Phi = \frac{1+\bar{\Phi}^2}{4 \bar{\Phi}} \,.
\end{equation}
The model functions \eqref{eq: ln model A}-\eqref{eq: ln model V} transform under eq.~\eqref{mapto} into
\begin{align}
\bar{\mathcal{A}}(\bar{\Phi}) &= \bar{\Phi}^{\scriptstyle\frac{1}{2\sqrt{6}}} \, \mathrm{e}^{-\frac{\sqrt{6}(\bar{\Phi}^4 -1)}{48 \bar{\Phi}^2}} \,,\\
\bar{\mathcal{B}}(\bar{\Phi}) &= \bar{\Phi}^{\scriptstyle\frac{1}{2\sqrt{6}}} \, \tfrac{(1-\bar{\Phi}^2)^2}{16 \bar{\Phi}^4} \, \mathrm{e}^{-\frac{\sqrt{6}(\bar{\Phi}^4 -1)}{48 \bar{\Phi}^2}}  \,, \\
\bar{\mathcal{V}}(\bar{\Phi}) &= \bar{\Phi}^{\scriptstyle\frac{1}{\sqrt{6}}} \, \mathrm{e}^{-\frac{\sqrt{6}(\bar{\Phi}^4 -1)}{24 \bar{\Phi}^2}} \left( \ln \tfrac{(1+\bar{\Phi}^2)^2}{16 \bar{\Phi}^2} \right)^2 \,,
\end{align}
and contrary to the previous form the functions $\bar{\mathcal{A}}(\bar{\Phi})$ and $\bar{\mathcal{B}}(\bar{\Phi})$ do not coincide any more. A direct integration of eq.~\eqref{I.3} yields the expressions of the invariant field now as
\begin{align}
\mathcal{I}_{\bar{\Phi}}^\Gamma &= \frac{1+\bar{\Phi}^2}{4\bar{\Phi}} \,, \\
\mathcal{I}_{\bar{\Phi}}^g &= \frac{1+\bar{\Phi}^4}{16\bar{\Phi}^2} + \frac{1}{8} \,,
\label{eq: ln model I_Phi_g_2}
\end{align} 
where the last term had to be added as an integration constant to maintain an explicit equivalence. By construction, we get the same invariant potentials \eqref{eq: ln model I_V_Gamma}, \eqref{eq: ln model I_V_g}, and PSRPs \eqref{eq: ln model epsilon}, \eqref{eq: ln model eta}. Note that if we had omitted the constant of integration in \eqref{eq: ln model I_Phi_g_2}, the proportionality of invariant fields \eqref{eq:IphimproptoIphim} would still hold, but the proportionality of invariant potentials \eqref{eq:IVmproptolnIphin} would not be completely obvious at first sight. Nevertheless, a direct calculation of the derivatives in the inflationary parameters \eqref{r}, \eqref{ns} would yield the same result with or without the integration constant. 


As a final comment, let us stress that in all the examples of this section the invariant PSRPs, and therefore the predictions for $r$, $n_s$ and $N$, coincide in the metric and Palatini cases, since the invariant potentials are proportional to each other and the overall factor cancels out. 
However, the amplitude of the scalar power spectrum \eqref{scalar.amplitude} depends explicitly on the invariant potential, and thus this observable will be sensitive to the difference in the actual normalization of the invariant potentials. The normalization can be crucial in satisfying the observational constraints, currently $A_s \simeq 2.1 \times 10^{-9}$ \cite{Akrami2018}. The metric and Palatini models will yield the same phenomenology also in this respect if a strict equivalence between the invariant potentials holds, not just a proportionality \eqref{eq:IVmproptoIVP}. Starting from exactly the same invariant actions, this is never the case. However, for the examples considered before, a change in the normalization of the model functions of the metric and Palatini action will have the final effect of generating exactly the same invariant potentials. For instance, for what concerns Example A, the invariant potential under metric and Palatini are the same when the non-minimal couplings in eq.~\eqref{eq:Coleman-Weinberg A} satisfy the following condition:
\begin{equation}
    \xi_\Gamma = \frac{\xi_g}{1+6\xi_g} \,,
\end{equation}
where $\xi_{g,\Gamma}$ are respectively the non-minimal coupling under the metric  and the Palatini formulation \cite{Racioppi2017,Racioppi2018}. A more general discussion about the generation of exactly equivalent invariant potentials is presented in the next section.

\section{When do different metric and Palatini actions yield the same observables?}
\label{sec:different_action}

As described in~\cite{Jarv2017}, equivalent inflationary theories are described by one invariant function\footnote{As already discussed in~\cite{Jarv2017}, the full gravitation equivalence needs to take into account also the invariant
$\mathcal{I}_m$, that describes the couplings to matter. Therefore, ensuring only the same invariant potential, the (p)reheating mechanism might still affect the
value of the observables and break the equivalence~\cite{Racioppi2017, Rubio:2019ypq}. On the other hand, we can see from \eqref{I.1} that,
by adjusting accordingly the function $\sigma$, we can easily obtain
theories where $\mathcal{I}_m$ is the same under both gravity
formulations, restoring the equivalence of observables also when
(p)reheating is considered.}: $\mathcal{I}_{\mathcal{V}} (\mathcal{I}_{{\Phi}})$. However, inflationary models can be produced by using three generating functions: $\mathcal{A}(\Phi)$, $\mathcal{B}(\Phi)$ and $\mathcal{V}(\Phi)$. Therefore, the a priori knowledge of $\mathcal{I}_{\mathcal{V}} (\mathcal{I}_{{\Phi}})$ allows us to derive only one constraint that $\mathcal{A}(\Phi)$, $\mathcal{B}(\Phi)$ and $\mathcal{V}(\Phi)$ have to satisfy, leaving two functions out of the three completely undetermined. Generally, we can express the invariant field $\mathcal{I}_\Phi$ as the inverse function of the invariant potential\footnote{The computation of $\mathcal{I}_{\mathcal{V}}^{-1}$ is quite delicate. In many cases $\mathcal{I}_{\mathcal{V}} (\mathcal{I}_{{\Phi}})$ is not a bijective (i.e. invertible) function, therefore $\mathcal{I}_{\mathcal{V}}^{-1}$ can be consistently identified only after a proper definition of the domain of $\mathcal{I}_{\mathcal{V}} (\mathcal{I}_{{\Phi}})$. For more details about this topic see \cite{Karamitsos:2019vor}, and also \cite{Jarv:2010zc,Jaerv2015a}. This is related to the possible occurrence of a singularity or exceptional point in the theory which could have implications for cosmology or in the presence of black holes (see e.g.~\cite{Briscese:2006xu, Bahamonde:2016wmz}). Nevertheless, here we restrict our attention to slow-roll inflation and smooth functions, expecting no singularities to arise.} in eq.~\eqref{I.2}
\begin{equation}
\mathcal{I}_{{\Phi}} = \Biggl( \frac{\mathcal{V}(\Phi)}{\mathcal{A}(\Phi)^2}\Biggr)^{-1} \equiv \mathcal{I}_{\mathcal{V}}^{-1}(\Phi) \, ,
\end{equation}
where the superscript ``$-1$" stands for inverse function. Using eq.~\eqref{I.3} we can write

\begin{equation}
\mathcal{I}_{\mathcal{V}}^{-1}(\Phi) = \int\!\mathrm{d}\Phi\,\sqrt{\frac{\mathcal{B}(\Phi)}{\mathcal{A}(\Phi)} 
		+  \frac{3}{2} \, \delta_{j\,\Gamma}  \, \left(\frac{\mathcal{A}'(\Phi)}{\mathcal{A}(\Phi)}\right)^2}\,,
		\label{eq:constraint}
\end{equation}
where the parameter $\delta_{j\,\Gamma}$ indicates the adopted gravity formulation. 
Therefore, given the invariant function $\mathcal{I}_{\mathcal{V}} (\mathcal{I}_{{\Phi}})$, eq.~\eqref{eq:constraint} is the constraint that $\mathcal{A}(\Phi)$, $\mathcal{B}(\Phi)$ and $\mathcal{V}(\Phi)$ must satisfy in order to create equivalent inflationary theories also among different gravity formulations. This means that (apart from pathological cases) we can randomly choose two functions among $\mathcal{A}(\Phi)$, $\mathcal{B}(\Phi)$ and $\mathcal{V}(\Phi)$. If the third one satisfies eq.~\eqref{eq:constraint}, then the correct $\mathcal{I}_{\mathcal{V}} (\mathcal{I}_{{\Phi}})$ is always generated. However, the solution  of the constraint \eqref{eq:constraint}  is strongly dependent on the initial choice of model functions, invariant potential and gravity formulation (metric or Palatini). Nevertheless, until the constraint \eqref{eq:constraint} is satisfied, the same invariant potential $\mathcal{I}_{\mathcal{V}} (\mathcal{I}_{{\Phi}})$ and therefore the same inflationary observables  (eqs. \eqref{r}-\eqref{number.of.efolds}) are generated, regardless of initial model functions and gravity formulation.

When  $\mathcal{A}(\Phi)$ and $\mathcal{V}(\Phi)$ are given, it is always possible to solve eq.~\eqref{eq:constraint} and obtain the corresponding value for the non-canonical kinetic function
\begin{equation}
 \mathcal{B}(\Phi) = \mathcal{A}(\Phi)
 \left\{\left[ \frac{d \, \mathcal{I}_{\mathcal{V}}^{-1}(\Phi) }{d \Phi}  \right]^2 -\frac{3}{2} \, \delta_{j\,\Gamma}  \, \left(\frac{\mathcal{A}'(\Phi)}{\mathcal{A}(\Phi)}\right)^2 \right\} \,,
 \label{eq:B}
\end{equation}
where $\delta_{j\,\Gamma}$ reflects the adopted gravity formulation (see eq.~\eqref{fl_fnide_teisendused}). 

Instead, if $\mathcal{A}(\Phi)$ and $\mathcal{B}(\Phi)$ are fixed, the constraint can be formally solved as
\begin{equation}
 \mathcal{V}(\Phi) =  \mathcal{A}(\Phi)^2 \, \mathcal{I}_{\mathcal{V}}\big(\mathcal{I}^{-1}_\mathcal{V}(\Phi)\big)   
 \label{eq:V}
\end{equation}
where $\mathcal{I}^{-1}_\mathcal{V}(\Phi)$ is given in eq.~\eqref{eq:constraint}. However, in this case, since the integral of an elementary function is not automatically elementary, choosing $\mathcal{A}(\Phi)$ and $\mathcal{B}(\Phi)$ as elementary functions of $\Phi$ does not always ensure that $\mathcal{V}(\Phi)$ is elementary as well. 

Finally, when $\mathcal{B}(\Phi)$ and $\mathcal{V}(\Phi)$ are chosen, the constraint eq.~\eqref{eq:constraint} becomes the following differential equation:
\begin{equation}
 \mathcal{A}(\Phi)
 \left\{\frac{3}{2} \, \delta_{j\,\Gamma}  \, \left(\frac{\mathcal{A}'(\Phi)}{\mathcal{A}(\Phi)}\right)^2  - \left[ \frac{d \, \mathcal{I}_{\mathcal{V}}^{-1}(\Phi) }{d \Phi}  \right]^2 \right\}+ \mathcal{B}(\Phi)  = 0 \,,
 \label{eq:A}
\end{equation}
to be solved in order to determine $\mathcal{A}(\Phi)$. 

Next, we present an example in order to better illustrate the different issues arising in each configuration.

\subsection{Example: Generalized Starobinsky invariant potential}

Let us consider the following invariant potential: 
\begin{equation}
\label{E.model.example}
\mathcal{I}_{\mathcal{V}}(\mathcal{I}_{\Phi}) 
= M^4 \left(1 - e^{-\sqrt{\frac{2 }{3 \alpha}} \mathcal{I}_{\Phi} } \right)^2\,,
\end{equation}
which generalizes the Starobinsky potential \cite{Ferrara2013b,Kallosh2013a,Galante2015}. The model is well-known too, therefore we just summarize briefly the main features. At the leading order in the invariant field value the slow-roll parameters are
\begin{eqnarray}
\epsilon &\approx& \frac{4}{3\alpha e^{2\sqrt{\frac{2 }{3 \alpha}} \mathcal{I}_{\Phi} }} \, , \\
\eta &\approx& \frac{-4}{3\alpha e^{\sqrt{\frac{2 }{3 \alpha}} \mathcal{I}_{\Phi} }} \, ,
\end{eqnarray}
while the number of $e$-folds is 
\begin{equation}
N \approx \frac{3\alpha}{4} e^{\sqrt{\frac{2}{3\alpha}}\mathcal{I}^*_\Phi} \, .
\end{equation}
Therefore, at the leading order in $N$ we get that
\begin{eqnarray}
n_s &\approx& 1 - \frac{2}{N} \, , \\
r &\approx& \frac{12 \alpha}{N^2} \, .
\end{eqnarray}
Planck data sets $\log_{10} \alpha < 1.3$ at $95\%$ CL \cite{Akrami2018}.

We can invert eq.~\eqref{E.model.example} to obtain 
\begin{equation}
\label{E.model.example.inverse} 
\mathcal{I}_{\mathcal{V}}^{-1}(\Phi) 
=-\sqrt{\frac{3 \alpha}{2}} \ln\left(1- \frac{1}{M^2} \sqrt{ \frac{\mathcal{V}(\Phi)}{\mathcal{A}(\Phi)^2}}\right) \,. 
\end{equation} 
Therefore, the constraint in eq.~\eqref{eq:constraint} becomes
\begin{align}
&\int\!\mathrm{d}\Phi\,\sqrt{\frac{\mathcal{B}(\Phi)}{\mathcal{A}(\Phi)} 
	+  \frac{3}{2} \, \delta_{j\,\Gamma}  \, \left(\frac{\mathcal{A}'(\Phi)}{\mathcal{A}(\Phi)}\right)^2} \nonumber  \\
&\qquad =-\sqrt{\frac{3 \alpha}{2}} \ln\left(1-\frac{1}{M^2}\sqrt{ \frac{\mathcal{V}(\Phi)}{\mathcal{A}(\Phi)^2}}\right) \,. \label{eq:constraint:example}
\end{align}
Let us consider now some case by case examples and see how the initial choice of the model functions affects the solving strategy of eq.~\eqref{eq:constraint:example}.

\subsubsection{$\mathcal{A}$ and $\mathcal{V}$ are fixed} 
Taking for instance the following natural inflation potential and non-minimal coupling to gravity
\begin{eqnarray}
\mathcal{A}(\Phi) &=& 1 + \xi \Phi^2 \label{eq:AV:A}\\ 
\mathcal{V}(\Phi) &=& M^4 \left( 1-\cos\left(\frac{\Phi}{\Phi_0}\right) \right)
\end{eqnarray}
we obtain the invariant potential in eq.~\eqref{E.model.example} if
\begin{equation}
\mathcal{B}(\Phi)_\Gamma =
\tfrac{3 \alpha}{4 \mathcal{A}(\Phi) \Phi_0^2}
\left(\tfrac{\mathcal{A}(\Phi) \cos \left(\frac{\Phi }{2 \Phi_0}\right)-4 \xi \Phi_0
   \Phi  \sin \left(\frac{\Phi }{2 \Phi_0}\right)}{\mathcal{A}(\Phi)-\sqrt{2} \sin \left(\frac{\Phi }{2 \Phi_0}\right)}\right)^2 
\end{equation}
in the Palatini case, with $\mathcal{A}(\Phi)$ given in eq.~\eqref{eq:AV:A} and
\begin{equation}
\mathcal{B}(\Phi)_g  = \mathcal{B}(\Phi)_\Gamma- 6 \frac{\xi^2 \Phi^2}{\mathcal{A}(\Phi)}  
\end{equation}
in the metric case.

\subsubsection{$\mathcal{A}$ and $\mathcal{B}$ are fixed} 
We consider now a non-minimally coupled scalar field  with a canonical kinetic term 
\begin{eqnarray}
\mathcal{A}(\Phi) &=& \frac{2}{3 \alpha} \Phi^2 \,, \\
\mathcal{B}(\Phi)&=& 1 \,.
\end{eqnarray}

Solving eq.~\eqref{eq:constraint}, we can see that we reproduce the invariant potential in eq.~\eqref{E.model.example} if the potential is
\begin{equation}
\mathcal{V}(\Phi)_{\Gamma} =
    \frac{4 M^4}{9 \alpha^2} \left(1 - \frac{\Phi_0}{\Phi}  \right)^2 \Phi^4
\end{equation}
in the Palatini case and 
\begin{equation}
\mathcal{V}(\Phi)_{g} =
    \frac{4 M^4}{9 \alpha^2} \left(1 - 
\left(\frac{\Phi_0}{\Phi} \right)^{\sqrt{\frac{4 +  \alpha}{\alpha}}}  \right)^2 \Phi^4
\end{equation}
in the metric case, where $\Phi_0$ is an integration constant.

\subsubsection{$\mathcal{B}$ and $\mathcal{V}$ are fixed} 
In this last example we consider a non-canonically normalized scalar and a quartic potential
\begin{eqnarray}
\mathcal{B}(\Phi) &=&\frac{6 (\alpha -1) \Phi ^2}{\mathcal{A}(\Phi )}\,, \label{eq:B:ex}\\ 
\mathcal{V}(\Phi) &=& M^4  \, \Phi ^4\,, \label{eq:V:ex}
\end{eqnarray}
with $\alpha>1$. We need to determine $\mathcal{A}(\Phi)$ by solving the differential equation in eq.~\eqref{eq:A}, where $\mathcal{I}_\mathcal{V}^{-1}$, $\mathcal{B}(\Phi)$ and $\mathcal{V}(\Phi)$ are given respectively in eqs.~\eqref{E.model.example.inverse}, \eqref{eq:B:ex} and \eqref{eq:V:ex}.
The specific choice in eq.~\eqref{eq:B:ex} allows us to solve such differential equation in both the metric and the Palatini cases. With a convenient choice of the integration constants, the corresponding solution is 
\begin{equation}
\mathcal{A}(\Phi)_{g} = 1+ \Phi^2 \, 
\end{equation}
in the metric case and
\begin{equation}
\mathcal{A}(\Phi)_{\Gamma} =\Phi^2 \left(1+  \, \Phi^{-2\sqrt{\frac{\alpha-1}{\alpha }}} \right) \, 
\label{eq:A:Gamma}
\end{equation}
in the Palatini case. The special 
value of $\alpha=1$ requires an additional comment. In this case the potential \eqref{E.model.example} becomes exactly the Starobinsky one and the non-canonical kinetic term \eqref{eq:B:ex} becomes identically zero in both the metric and the Palatini formulations. For the first case this not a problem because it coincides with the formulation of the Starobinsky model via the auxiliary field in the Jordan frame. On the other hand, as discussed in Section \ref{sec:action}, in the Palatini formulation the invariant field $\mathcal{I}_\Phi$ is not dynamical and the problem does not have a solution. However, it is still possible to reproduce the potential \eqref{E.model.example} from \eqref{eq:V:ex} in the Palatini formulation by relaxing the condition \eqref{eq:B:ex}. For instance, choosing 
\begin{equation}
\mathcal{B}(\Phi)_{\Gamma} = \frac{\alpha \Phi ^2}{\mathcal{A}(\Phi )}  \, ,
\end{equation}
we would get 
\begin{equation}
\mathcal{A}(\Phi)_{\Gamma} = \Phi ^2  \left( 1 + \Phi ^{-\sqrt{\frac{2}{3}}} \right) \, .
\end{equation}

\section{Summary and conclusions}
\label{sec:conclusions}

In the present paper we studied the slow-roll parameters and inflationary observables in the framework of scalar-tensor theories of gravity in the metric and Palatini formulations. The model functions $\mathcal{A}(\Phi)$, $\mathcal{B}(\Phi)$, $\mathcal{V}(\Phi)$ allow us to construct quantities, which are invariant under a conformal transformation of the metric and behave as scalar functions under the scalar field redefinition. Using this frame invariant 
approach we expressed the slow-roll parameters $\epsilon$, $\eta$, 
as well as the inflationary observable quantities $n_{s}$, $r$, $A_{s}$, and explained in detail how to compute them in the case of different model functions. 

Next, in the main part of the paper, we clarified what conditions must be met for the metric and Palatini formalisms to give the same observable quantities. Due to the fact that most of the observable quantities are independent of the overall normalization factor, we concluded that it is sufficient for the invariant potentials in both formulations to be proportional to each other in order to obtain equal predictions for $r$, $n_s$ and $N$ (but not $A_s$) in both formulations. We illustrated this general statement by two specific examples. After that, starting from the same invariant potential, we showed how by fixing two out of the three model functions we can straightforwardly obtain the third. We demonstrated the different possibilities by considering as an example an invariant potential of the Starobinsky form. One then  
sees how seemingly different models of inflation can give the same values of the observed parameters. 

A deeper case-by-case study may unveil other configurations where the same model functions, but with different values of the free parameters, share the same invariant potential and therefore give the same values for observables. 
The framework described here provides a tool that enables to easily check different models against observations, as well as to reconstruct variations of models with a given phenomenology.

If the next generation satellites (LITEBIRD \cite{Matsumura2016}, PIXIE \cite{Kogut_2011}, PICO \cite{Sutin:2018onu}) will be launched and after data will be collected, the available parameter space will be even more constrained, leaving us with a reduced set of allowed invariant potentials and more indications about which gravity formulation satisfies additional criteria like \emph{elegance, simplicity} or \emph{minimality}.

\begin{acknowledgments}
	This work was supported by the Estonian Research Council grants MOBJD381, MOBTT5, MOBTT86, PRG356 and by the EU through the European Regional Development Fund CoE program TK133 ``The Dark Side of the Universe." A. Kozak was a beneficiary of the "International scholarship exchange of PhD candidates and academic staff" programme (PROM) organized by Polish National Agency for Academic Exchange (NAWA). The research of A. Lykkas is co-financed by Greece and the European Union (European Social Fund - ESF) through the Operational Programme “Human Resources Development, Education and Lifelong Learning” in the context of the project “Strengthening Human Resources Research Potential via Doctorate Research” (MIS-5000432), implemented by the State Scholarships Foundation (IKY). 
\end{acknowledgments}

\appendix
\begin{widetext}
\section{Appendix} \label{appendix}
The most general action for a class of Palatini scalar-tensor theories of gravity featuring non-metricity vectors entering the action functional in a linear way can be written as follows~\cite{Kozak:2018vlp, Borowiec:2020lfx}:
\begin{equation}
\label{actionGP}
\mathcal{S} = \int\!\mathrm{d}^4x\,\sqrt{-g} \left\{\frac{1}{2}\mathcal{A}(\Phi)R(g,\Gamma) - \frac{1}{2}\mathcal{B}(\Phi)(\nabla\Phi)^2 -\mathcal{V}(\Phi) -\mathcal{C}_1(\Phi) Q^\mu\nabla_\mu\Phi - \mathcal{C}_2(\Phi)\bar{Q}^\mu\nabla_\mu\Phi \right\} + \mathcal{S}_m[\mathrm{e}^{2\sigma(\Phi)} g_{\mu\nu}, \chi_{m}].
\end{equation}
The action contains three independent variables: metric tensor, affine connection, and scalar field. It also features six arbitrary functions of the scalar field: $\{\mathcal{A},\mathcal{B},\mathcal{C}_1,\mathcal{C}_2,\mathcal{V},\sigma\}$, providing, together with the dynamical variables, the so-called ``frame'' for the action \eqref{actionGP}. 
The vectors $Q^\mu$ and $\bar{Q}^\mu$ are defined as
\begin{subequations}
		\begin{align}
	& Q^\mu=g^{\mu\nu}g^{\alpha\beta}\nabla^\Gamma_\nu g_{\alpha\beta}=g^{\mu\nu}g^{\alpha\beta}Q_{\nu\alpha\beta}, \\
	& \bar{Q}^\mu=-g^{\mu\nu}g^{\alpha\beta}\nabla^\Gamma_\alpha g_{\nu\beta}=-g^{\mu\nu}g^{\alpha\beta}Q_{\alpha\nu\beta}.
		\end{align}
	\end{subequations}
The $\nabla^\Gamma$ is defined with respect to the independent connection, therefore the covariant derivative of the metric will not vanish in general. 

In the Palatini approach, the metric tensor is fundamentally independent of the connection. When we use the Weyl (or conformal) transformation of the metric, the connection remains unchanged. We might use this freedom and postulate additional transformations of the connection preserving the light cones. We introduce the following transformation formulae for the dynamical variables entering the action functional:
\begin{subequations}
			\begin{align}
			 g_{\mu\nu}&=\mathrm{e}^{2\bar{\gamma}_1(\bar{\Phi})}\bar{g}_{\mu\nu}, \label{e1} \\ 
			 \Gamma^\alpha_{\mu\nu}&=\bar{\Gamma}^\alpha_{\mu\nu}+2 \delta^\alpha{}_{(\mu}\partial_{\nu)}\bar{\gamma}_2(\bar{\Phi})-\bar{g}_{\mu\nu}\bar{g}^{\alpha\beta}\partial_\beta\bar{\gamma}_3(\bar{\Phi}), \label{e2}\\
			 \Phi&=\bar{f}(\bar{\Phi}). \label{e3}
			\end{align}
		\end{subequations}
The transformations are governed by three smooth functions of the scalar field $(\gamma_1, \gamma_2, \gamma_3)$, and are accompanied  by a redefinition of the scalar field. 
The transformations \eqref{e1})-\eqref{e3}) are invertible
\begin{subequations}
			\begin{align}
			 \bar{g}_{\mu\nu}&=\mathrm{e}^{2\gamma_1(\Phi)}g_{\mu\nu}, \\ 
			 \bar{\Gamma}^\alpha_{\mu\nu}&=\Gamma^\alpha_{\mu\nu}+2 \delta^\alpha{}_{(\mu}\partial_{\nu)}\gamma_2(\Phi)-g_{\mu\nu}g^{\alpha\beta}\partial_\beta\gamma_3(\Phi), \\
			 \bar{\Phi}&=f(\Phi), 
			\end{align}
\end{subequations}
and the relations between the gamma functions and the diffeomorphism of the scalar field are given by:
\begin{subequations}
			\begin{align}
			&\bar{\gamma}_i^{}=-\gamma_i \circ f, \label{inverse1}\\
			 & \bar{f} =f^{-1}.\label{inverse2}
			\end{align}
\end{subequations}
The action \eqref{actionGP} turns out to be form-invariant under the action of transformations \eqref{e1}-\eqref{e3}, which means that solutions to the field equations obtained in one frame are mapped into corresponding solutions in another frame, assuming that the six functions of the scalar field $\{\mathcal{A},\mathcal{B},\mathcal{C}_1,\mathcal{C}_2,\mathcal{V},\alpha\}$ change in the following way: 
	\begin{subequations}
		\begin{align}
		\mathcal{\bar{A}}(\bar{\Phi})&=\mathrm{e}^{2\bar{\gamma}_1(\bar{\Phi})}\mathcal{A}(\bar{f}(\bar{\Phi})), \label{t1}\\
		\mathcal{\bar{B}}(\bar{\Phi})&=\mathrm{e}^{2\bar{\gamma}_1(\bar{\Phi})}\Bigg[\mathcal{B}(\bar{f}(\bar{\Phi}))(\bar{f}'(\bar{\Phi}))^2\nonumber\\
		&\qquad+\bar{f}'(\bar{\Phi})\Big(\mathcal{C}_1(\bar{f}(\bar{\Phi}))(8\bar{\gamma}'_1(\bar{\Phi})-10\bar{\gamma}'_2(\bar{\Phi})+2\bar{\gamma}'_3(\bar{\Phi}))-\mathcal{C}_2(\bar{f}(\bar{\Phi}))(2\bar{\gamma}'_1(\bar{\Phi})-7\bar{\gamma}'_2(\bar{\Phi})+5\bar{\gamma}'_3(\bar{\Phi}))\Big)\nonumber\\
			&\qquad\qquad+3\Big(4\mathcal{A}(\bar{f}(\bar{\Phi}))\bar{\gamma}'_2(\bar{\Phi})\bar{\gamma}'_3(\bar{\Phi})-\mathcal{A}(\bar{f}(\bar{\Phi}))\left(\bar{\gamma}'_2(\bar{\Phi})\right)^2-\mathcal{A}(\bar{f}(\bar{\Phi}))\left(\bar{\gamma}'_3(\bar{\Phi})\right)^2\nonumber\\
			&\qquad\qquad\qquad+\frac{\mathrm{d}\mathcal{A}(\bar{f}(\bar{\Phi}))}{\mathrm{f}\bar{\Phi}}(\bar{\gamma}'_2(\bar{\Phi})+\bar{\gamma}'_3(\bar{\Phi}))-2\mathcal{A}(\bar{f}(\bar{\Phi}))\bar{\gamma}'_1(\bar{\Phi})(\bar{\gamma}'_2(\bar{\Phi})+\bar{\gamma}'_3(\bar{\Phi}))\Big)\Bigg], \\
		\mathcal{\bar{C}}_1(\bar{\Phi})&=\mathrm{e}^{2\bar{\gamma}_1(\bar{\Phi})}\Big[\bar{f}'(\bar{\Phi})\mathcal{C}_1(\bar{f}(\bar{\Phi}))-\mathcal{A}(\bar{f}(\bar{\Phi}))\left(\frac{3}{2}\bar{\gamma}'_2(\bar{\Phi})+\frac{1}{2}\bar{\gamma}'_3(\bar{\Phi})\right)\Big],\\
		\mathcal{\bar{C}}_2(\bar{\Phi})&=\mathrm{e}^{2\bar{\gamma}_1(\bar{\Phi})}\Big[\bar{f}'(\bar{\Phi})\mathcal{C}_2(\bar{f}(\bar{\Phi}))-\mathcal{A}(\bar{f}(\bar{\Phi}))\left(3\bar{\gamma}'_2(\bar{\Phi})-\bar{\gamma}'_3(\bar{\Phi})\right)\Big],\\
		\mathcal{\bar{V}}(\bar{\Phi})&=\mathrm{e}^{4\bar{\gamma}_1(\bar{\Phi})}\mathcal{V}(\bar{f}(\bar{\Phi})), \\
		\bar{\sigma}(\bar{\Phi})&=\sigma(\bar{f}(\bar{\Phi}))+\bar{\gamma}_1(\bar{\Phi}).\label{t6}
		\end{align} \label{transformations}
	\end{subequations}

It is always possible to choose the functions $(\gamma_2, \gamma_3)$ in such a way that the functions $\mathcal{C}_1$ and $\mathcal{C}_2$ vanish. Indeed, one must take
\begin{subequations}
\begin{align}
& \bar{\gamma}'_2(\Phi) = \frac{-2\mathcal{C}_1(\Phi) -\mathcal{C}_2(\Phi)}{6\mathcal{A}(\Phi)}, \\
& \bar{\gamma}'_3(\Phi) = \frac{-2\mathcal{C}_1(\Phi) + \mathcal{C}_2(\Phi)}{2\mathcal{A}(\Phi)}.
\end{align}
\end{subequations}
Such a choice will transform the action \eqref{actionGP} to the following one:
\begin{equation}
\mathcal{S} =\int\!\mathrm{d}^4x\,\sqrt{-g} \Big\{\frac{1}{2}\mathcal{A}(\Phi)R(g,\bar{\Gamma}) - \frac{1}{2}\mathcal{\bar{B}}(\Phi)(\nabla\Phi)^2 -\mathcal{V}(\Phi)\Big\}  + \mathcal{S}_m[\mathrm{e}^{2\sigma(\Phi)} g_{\mu\nu}, \chi_{m}],
\end{equation}
where
\begin{equation}
\mathcal{\bar{B}}(\Phi) =  \mathcal{B}(\Phi) + \frac{\mathcal{A}'(\Phi)\big(\mathcal{C}_2(\Phi)-4 \mathcal{C}_1(\Phi)\big)}{\mathcal{A}(\Phi)} +\frac{ 11\mathcal{C}^2_2(\Phi) - 4\mathcal{C}^2_1(\Phi) - 16\mathcal{C}_1(\Phi)\mathcal{C}_2(\Phi)}{6\mathcal{A}(\Phi)},
\end{equation}
which justifies the choice of the initial action \eqref{eq:action} without the $\mathcal{C}_i$ functions. 
\end{widetext}

\bibliography{References}

\begin{thebibliography}{89}%
\makeatletter
\providecommand \@ifxundefined [1]{%
 \@ifx{#1\undefined}
}%
\providecommand \@ifnum [1]{%
 \ifnum #1\expandafter \@firstoftwo
 \else \expandafter \@secondoftwo
 \fi
}%
\providecommand \@ifx [1]{%
 \ifx #1\expandafter \@firstoftwo
 \else \expandafter \@secondoftwo
 \fi
}%
\providecommand \natexlab [1]{#1}%
\providecommand \enquote  [1]{``#1''}%
\providecommand \bibnamefont  [1]{#1}%
\providecommand \bibfnamefont [1]{#1}%
\providecommand \citenamefont [1]{#1}%
\providecommand \href@noop [0]{\@secondoftwo}%
\providecommand \href [0]{\begingroup \@sanitize@url \@href}%
\providecommand \@href[1]{\@@startlink{#1}\@@href}%
\providecommand \@@href[1]{\endgroup#1\@@endlink}%
\providecommand \@sanitize@url [0]{\catcode `\\12\catcode `\$12\catcode
  `\&12\catcode `\#12\catcode `\^12\catcode `\_12\catcode `\%12\relax}%
\providecommand \@@startlink[1]{}%
\providecommand \@@endlink[0]{}%
\providecommand \url  [0]{\begingroup\@sanitize@url \@url }%
\providecommand \@url [1]{\endgroup\@href {#1}{\urlprefix }}%
\providecommand \urlprefix  [0]{URL }%
\providecommand \Eprint [0]{\href }%
\providecommand \doibase [0]{http://dx.doi.org/}%
\providecommand \selectlanguage [0]{\@gobble}%
\providecommand \bibinfo  [0]{\@secondoftwo}%
\providecommand \bibfield  [0]{\@secondoftwo}%
\providecommand \translation [1]{[#1]}%
\providecommand \BibitemOpen [0]{}%
\providecommand \bibitemStop [0]{}%
\providecommand \bibitemNoStop [0]{.\EOS\space}%
\providecommand \EOS [0]{\spacefactor3000\relax}%
\providecommand \BibitemShut  [1]{\csname bibitem#1\endcsname}%
\let\auto@bib@innerbib\@empty
\bibitem [{\citenamefont {Akrami}\ \emph {et~al.}(2018)\citenamefont {Akrami}
  \emph {et~al.}}]{Akrami2018}%
  \BibitemOpen
  \bibfield  {author} {\bibinfo {author} {\bibfnamefont {Y.}~\bibnamefont
  {Akrami}} \emph {et~al.} (\bibinfo {collaboration} {Planck}),\ }\bibfield
  {title} {\enquote {\bibinfo {title} {{Planck 2018 results. X. Constraints on
  inflation}},}\ }\href@noop {} {\  (\bibinfo {year} {2018})},\ \Eprint
  {http://arxiv.org/abs/1807.06211} {arXiv:1807.06211 [astro-ph.CO]}
  \BibitemShut {NoStop}%
\bibitem [{\citenamefont {Starobinsky}(1980)}]{Starobinsky1980}%
  \BibitemOpen
  \bibfield  {author} {\bibinfo {author} {\bibfnamefont {Alexei~A.}\
  \bibnamefont {Starobinsky}},\ }\bibfield  {title} {\enquote {\bibinfo {title}
  {{A New Type of Isotropic Cosmological Models Without Singularity}},}\ }\href
  {\doibase 10.1016/0370-2693(80)90670-X} {\bibfield  {journal} {\bibinfo
  {journal} {Phys. Lett.}\ }\textbf {\bibinfo {volume} {91B}},\ \bibinfo
  {pages} {99--102} (\bibinfo {year} {1980})}\BibitemShut {NoStop}%
\bibitem [{\citenamefont {Bezrukov}\ and\ \citenamefont
  {Shaposhnikov}(2008)}]{Bezrukov2008}%
  \BibitemOpen
  \bibfield  {author} {\bibinfo {author} {\bibfnamefont {Fedor~L.}\
  \bibnamefont {Bezrukov}}\ and\ \bibinfo {author} {\bibfnamefont {Mikhail}\
  \bibnamefont {Shaposhnikov}},\ }\bibfield  {title} {\enquote {\bibinfo
  {title} {{The Standard Model Higgs boson as the inflaton}},}\ }\href
  {\doibase 10.1016/j.physletb.2007.11.072} {\bibfield  {journal} {\bibinfo
  {journal} {Phys. Lett.}\ }\textbf {\bibinfo {volume} {B659}},\ \bibinfo
  {pages} {703--706} (\bibinfo {year} {2008})},\ \Eprint
  {http://arxiv.org/abs/0710.3755} {arXiv:0710.3755 [hep-th]} \BibitemShut
  {NoStop}%
\bibitem [{\citenamefont {Bezrukov}\ \emph {et~al.}(2009)\citenamefont
  {Bezrukov}, \citenamefont {Magnin},\ and\ \citenamefont
  {Shaposhnikov}}]{Bezrukov2009a}%
  \BibitemOpen
  \bibfield  {author} {\bibinfo {author} {\bibfnamefont {Fedor~L.}\
  \bibnamefont {Bezrukov}}, \bibinfo {author} {\bibfnamefont {Amaury}\
  \bibnamefont {Magnin}}, \ and\ \bibinfo {author} {\bibfnamefont {Mikhail}\
  \bibnamefont {Shaposhnikov}},\ }\bibfield  {title} {\enquote {\bibinfo
  {title} {{Standard Model Higgs boson mass from inflation}},}\ }\href
  {\doibase 10.1016/j.physletb.2009.03.035} {\bibfield  {journal} {\bibinfo
  {journal} {Phys. Lett.}\ }\textbf {\bibinfo {volume} {B675}},\ \bibinfo
  {pages} {88--92} (\bibinfo {year} {2009})},\ \Eprint
  {http://arxiv.org/abs/0812.4950} {arXiv:0812.4950 [hep-ph]} \BibitemShut
  {NoStop}%
\bibitem [{\citenamefont {Bezrukov}\ and\ \citenamefont
  {Shaposhnikov}(2009)}]{Bezrukov2009}%
  \BibitemOpen
  \bibfield  {author} {\bibinfo {author} {\bibfnamefont {F.}~\bibnamefont
  {Bezrukov}}\ and\ \bibinfo {author} {\bibfnamefont {M.}~\bibnamefont
  {Shaposhnikov}},\ }\bibfield  {title} {\enquote {\bibinfo {title} {{Standard
  Model Higgs boson mass from inflation: Two loop analysis}},}\ }\href
  {\doibase 10.1088/1126-6708/2009/07/089} {\bibfield  {journal} {\bibinfo
  {journal} {JHEP}\ }\textbf {\bibinfo {volume} {07}},\ \bibinfo {pages} {089}
  (\bibinfo {year} {2009})},\ \Eprint {http://arxiv.org/abs/0904.1537}
  {arXiv:0904.1537 [hep-ph]} \BibitemShut {NoStop}%
\bibitem [{\citenamefont {Bezrukov}\ \emph {et~al.}(2011)\citenamefont
  {Bezrukov}, \citenamefont {Magnin}, \citenamefont {Shaposhnikov},\ and\
  \citenamefont {Sibiryakov}}]{Bezrukov2011}%
  \BibitemOpen
  \bibfield  {author} {\bibinfo {author} {\bibfnamefont {F.}~\bibnamefont
  {Bezrukov}}, \bibinfo {author} {\bibfnamefont {A.}~\bibnamefont {Magnin}},
  \bibinfo {author} {\bibfnamefont {M.}~\bibnamefont {Shaposhnikov}}, \ and\
  \bibinfo {author} {\bibfnamefont {S.}~\bibnamefont {Sibiryakov}},\ }\bibfield
   {title} {\enquote {\bibinfo {title} {{Higgs inflation: consistency and
  generalisations}},}\ }\href {\doibase 10.1007/JHEP01(2011)016} {\bibfield
  {journal} {\bibinfo  {journal} {JHEP}\ }\textbf {\bibinfo {volume} {01}},\
  \bibinfo {pages} {016} (\bibinfo {year} {2011})},\ \Eprint
  {http://arxiv.org/abs/1008.5157} {arXiv:1008.5157 [hep-ph]} \BibitemShut
  {NoStop}%
\bibitem [{\citenamefont {Quiros}\ \emph {et~al.}(2013)\citenamefont {Quiros},
  \citenamefont {Garcia-Salcedo}, \citenamefont {Madriz~Aguilar},\ and\
  \citenamefont {Matos}}]{Quiros2013}%
  \BibitemOpen
  \bibfield  {author} {\bibinfo {author} {\bibfnamefont {Israel}\ \bibnamefont
  {Quiros}}, \bibinfo {author} {\bibfnamefont {Ricardo}\ \bibnamefont
  {Garcia-Salcedo}}, \bibinfo {author} {\bibfnamefont {Jose~Edgar}\
  \bibnamefont {Madriz~Aguilar}}, \ and\ \bibinfo {author} {\bibfnamefont
  {Tonatiuh}\ \bibnamefont {Matos}},\ }\bibfield  {title} {\enquote {\bibinfo
  {title} {{The conformal transformation's controversy: what are we
  missing?}}}\ }\href {\doibase 10.1007/s10714-012-1484-7} {\bibfield
  {journal} {\bibinfo  {journal} {Gen. Rel. Grav.}\ }\textbf {\bibinfo {volume}
  {45}},\ \bibinfo {pages} {489--518} (\bibinfo {year} {2013})},\ \Eprint
  {http://arxiv.org/abs/1108.5857} {arXiv:1108.5857 [gr-qc]} \BibitemShut
  {NoStop}%
\bibitem [{\citenamefont {Chiba}\ and\ \citenamefont
  {Yamaguchi}(2013)}]{Chiba2013a}%
  \BibitemOpen
  \bibfield  {author} {\bibinfo {author} {\bibfnamefont {Takeshi}\ \bibnamefont
  {Chiba}}\ and\ \bibinfo {author} {\bibfnamefont {Masahide}\ \bibnamefont
  {Yamaguchi}},\ }\bibfield  {title} {\enquote {\bibinfo {title}
  {{Conformal-Frame (In)dependence of Cosmological Observations in
  Scalar-Tensor Theory}},}\ }\href {\doibase 10.1088/1475-7516/2013/10/040}
  {\bibfield  {journal} {\bibinfo  {journal} {JCAP}\ }\textbf {\bibinfo
  {volume} {1310}},\ \bibinfo {pages} {040} (\bibinfo {year} {2013})},\ \Eprint
  {http://arxiv.org/abs/1308.1142} {arXiv:1308.1142 [gr-qc]} \BibitemShut
  {NoStop}%
\bibitem [{\citenamefont {Kamenshchik}\ and\ \citenamefont
  {Steinwachs}(2015)}]{Kamenshchik:2014waa}%
  \BibitemOpen
  \bibfield  {author} {\bibinfo {author} {\bibfnamefont {Alexander~Yu.}\
  \bibnamefont {Kamenshchik}}\ and\ \bibinfo {author} {\bibfnamefont
  {Christian~F.}\ \bibnamefont {Steinwachs}},\ }\bibfield  {title} {\enquote
  {\bibinfo {title} {{Question of quantum equivalence between Jordan frame and
  Einstein frame}},}\ }\href {\doibase 10.1103/PhysRevD.91.084033} {\bibfield
  {journal} {\bibinfo  {journal} {Phys. Rev. D}\ }\textbf {\bibinfo {volume}
  {91}},\ \bibinfo {pages} {084033} (\bibinfo {year} {2015})},\ \Eprint
  {http://arxiv.org/abs/1408.5769} {arXiv:1408.5769 [gr-qc]} \BibitemShut
  {NoStop}%
\bibitem [{\citenamefont {Domènech}\ and\ \citenamefont
  {Sasaki}(2015)}]{Domenech2015a}%
  \BibitemOpen
  \bibfield  {author} {\bibinfo {author} {\bibfnamefont {Guillem}\ \bibnamefont
  {Domènech}}\ and\ \bibinfo {author} {\bibfnamefont {Misao}\ \bibnamefont
  {Sasaki}},\ }\bibfield  {title} {\enquote {\bibinfo {title} {{Conformal Frame
  Dependence of Inflation}},}\ }\href {\doibase 10.1088/1475-7516/2015/04/022}
  {\bibfield  {journal} {\bibinfo  {journal} {JCAP}\ }\textbf {\bibinfo
  {volume} {1504}},\ \bibinfo {pages} {022} (\bibinfo {year} {2015})},\ \Eprint
  {http://arxiv.org/abs/1501.07699} {arXiv:1501.07699 [gr-qc]} \BibitemShut
  {NoStop}%
\bibitem [{\citenamefont {Herrero-Valea}(2016)}]{Herrero-Valea2016}%
  \BibitemOpen
  \bibfield  {author} {\bibinfo {author} {\bibfnamefont {Mario}\ \bibnamefont
  {Herrero-Valea}},\ }\bibfield  {title} {\enquote {\bibinfo {title}
  {{Anomalies, equivalence and renormalization of cosmological frames}},}\
  }\href {\doibase 10.1103/PhysRevD.93.105038} {\bibfield  {journal} {\bibinfo
  {journal} {Phys. Rev.}\ }\textbf {\bibinfo {volume} {D93}},\ \bibinfo {pages}
  {105038} (\bibinfo {year} {2016})},\ \Eprint
  {http://arxiv.org/abs/1602.06962} {arXiv:1602.06962 [hep-th]} \BibitemShut
  {NoStop}%
\bibitem [{\citenamefont {Burns}\ \emph {et~al.}(2016)\citenamefont {Burns},
  \citenamefont {Karamitsos},\ and\ \citenamefont {Pilaftsis}}]{Burns2016}%
  \BibitemOpen
  \bibfield  {author} {\bibinfo {author} {\bibfnamefont {Daniel}\ \bibnamefont
  {Burns}}, \bibinfo {author} {\bibfnamefont {Sotirios}\ \bibnamefont
  {Karamitsos}}, \ and\ \bibinfo {author} {\bibfnamefont {Apostolos}\
  \bibnamefont {Pilaftsis}},\ }\bibfield  {title} {\enquote {\bibinfo {title}
  {{Frame-Covariant Formulation of Inflation in Scalar-Curvature Theories}},}\
  }\href {\doibase 10.1016/j.nuclphysb.2016.04.036} {\bibfield  {journal}
  {\bibinfo  {journal} {Nucl. Phys.}\ }\textbf {\bibinfo {volume} {B907}},\
  \bibinfo {pages} {785--819} (\bibinfo {year} {2016})},\ \Eprint
  {http://arxiv.org/abs/1603.03730} {arXiv:1603.03730 [hep-ph]} \BibitemShut
  {NoStop}%
\bibitem [{\citenamefont {Brooker}\ \emph {et~al.}(2016)\citenamefont
  {Brooker}, \citenamefont {Odintsov},\ and\ \citenamefont
  {Woodard}}]{Brooker2016}%
  \BibitemOpen
  \bibfield  {author} {\bibinfo {author} {\bibfnamefont {D.~J.}\ \bibnamefont
  {Brooker}}, \bibinfo {author} {\bibfnamefont {S.~D.}\ \bibnamefont
  {Odintsov}}, \ and\ \bibinfo {author} {\bibfnamefont {R.~P.}\ \bibnamefont
  {Woodard}},\ }\bibfield  {title} {\enquote {\bibinfo {title} {{Precision
  predictions for the primordial power spectra from $f(R)$ models of
  inflation}},}\ }\href {\doibase 10.1016/j.nuclphysb.2016.08.010} {\bibfield
  {journal} {\bibinfo  {journal} {Nucl. Phys.}\ }\textbf {\bibinfo {volume}
  {B911}},\ \bibinfo {pages} {318--337} (\bibinfo {year} {2016})},\ \Eprint
  {http://arxiv.org/abs/1606.05879} {arXiv:1606.05879 [gr-qc]} \BibitemShut
  {NoStop}%
\bibitem [{\citenamefont {Bhattacharya}\ and\ \citenamefont
  {Majhi}(2017)}]{Bhattacharya2017a}%
  \BibitemOpen
  \bibfield  {author} {\bibinfo {author} {\bibfnamefont {Krishnakanta}\
  \bibnamefont {Bhattacharya}}\ and\ \bibinfo {author} {\bibfnamefont
  {Bibhas~Ranjan}\ \bibnamefont {Majhi}},\ }\bibfield  {title} {\enquote
  {\bibinfo {title} {{Fresh look at the scalar-tensor theory of gravity in
  Jordan and Einstein frames from undiscussed standpoints}},}\ }\href {\doibase
  10.1103/PhysRevD.95.064026} {\bibfield  {journal} {\bibinfo  {journal} {Phys.
  Rev.}\ }\textbf {\bibinfo {volume} {D95}},\ \bibinfo {pages} {064026}
  (\bibinfo {year} {2017})},\ \Eprint {http://arxiv.org/abs/1702.07166}
  {arXiv:1702.07166 [gr-qc]} \BibitemShut {NoStop}%
\bibitem [{\citenamefont {Pandey}\ and\ \citenamefont
  {Banerjee}(2017)}]{Pandey2017}%
  \BibitemOpen
  \bibfield  {author} {\bibinfo {author} {\bibfnamefont {Sachin}\ \bibnamefont
  {Pandey}}\ and\ \bibinfo {author} {\bibfnamefont {Narayan}\ \bibnamefont
  {Banerjee}},\ }\bibfield  {title} {\enquote {\bibinfo {title} {{Equivalence
  of Jordan and Einstein frames at the quantum level}},}\ }\href {\doibase
  10.1140/epjp/i2017-11385-0} {\bibfield  {journal} {\bibinfo  {journal} {Eur.
  Phys. J. Plus}\ }\textbf {\bibinfo {volume} {132}},\ \bibinfo {pages} {107}
  (\bibinfo {year} {2017})},\ \Eprint {http://arxiv.org/abs/1610.00584}
  {arXiv:1610.00584 [gr-qc]} \BibitemShut {NoStop}%
\bibitem [{\citenamefont {Bahamonde}\ \emph {et~al.}(2017)\citenamefont
  {Bahamonde}, \citenamefont {Odintsov}, \citenamefont {Oikonomou},\ and\
  \citenamefont {Tretyakov}}]{Bahamonde2017}%
  \BibitemOpen
  \bibfield  {author} {\bibinfo {author} {\bibfnamefont {Sebastian}\
  \bibnamefont {Bahamonde}}, \bibinfo {author} {\bibfnamefont {Sergei~D.}\
  \bibnamefont {Odintsov}}, \bibinfo {author} {\bibfnamefont {V.~K.}\
  \bibnamefont {Oikonomou}}, \ and\ \bibinfo {author} {\bibfnamefont {Petr~V.}\
  \bibnamefont {Tretyakov}},\ }\bibfield  {title} {\enquote {\bibinfo {title}
  {{Deceleration versus acceleration universe in different frames of $F(R)$
  gravity}},}\ }\href {\doibase 10.1016/j.physletb.2017.01.012} {\bibfield
  {journal} {\bibinfo  {journal} {Phys. Lett.}\ }\textbf {\bibinfo {volume}
  {B766}},\ \bibinfo {pages} {225--230} (\bibinfo {year} {2017})},\ \Eprint
  {http://arxiv.org/abs/1701.02381} {arXiv:1701.02381 [gr-qc]} \BibitemShut
  {NoStop}%
\bibitem [{\citenamefont {Karamitsos}\ and\ \citenamefont
  {Pilaftsis}(2017)}]{Karamitsos2017}%
  \BibitemOpen
  \bibfield  {author} {\bibinfo {author} {\bibfnamefont {Sotirios}\
  \bibnamefont {Karamitsos}}\ and\ \bibinfo {author} {\bibfnamefont
  {Apostolos}\ \bibnamefont {Pilaftsis}},\ }\bibfield  {title} {\enquote
  {\bibinfo {title} {{Frame Covariant Nonminimal Multifield Inflation}},}\
  }\href {\doibase 10.1016/j.nuclphysb.2017.12.015} {\bibfield  {journal}
  {\bibinfo  {journal} {Nucl. Phys.}\ }\textbf {\bibinfo {volume} {B927}},\
  \bibinfo {pages} {219--254} (\bibinfo {year} {2017})},\ \Eprint
  {http://arxiv.org/abs/1706.07011} {arXiv:1706.07011 [hep-ph]} \BibitemShut
  {NoStop}%
\bibitem [{\citenamefont {Ruf}\ and\ \citenamefont
  {Steinwachs}(2018)}]{Ruf2018}%
  \BibitemOpen
  \bibfield  {author} {\bibinfo {author} {\bibfnamefont {Michael~S.}\
  \bibnamefont {Ruf}}\ and\ \bibinfo {author} {\bibfnamefont {Christian~F.}\
  \bibnamefont {Steinwachs}},\ }\bibfield  {title} {\enquote {\bibinfo {title}
  {{Quantum equivalence of $f(R)$ gravity and scalar-tensor theories}},}\
  }\href {\doibase 10.1103/PhysRevD.97.044050} {\bibfield  {journal} {\bibinfo
  {journal} {Phys. Rev.}\ }\textbf {\bibinfo {volume} {D97}},\ \bibinfo {pages}
  {044050} (\bibinfo {year} {2018})},\ \Eprint
  {http://arxiv.org/abs/1711.07486} {arXiv:1711.07486 [gr-qc]} \BibitemShut
  {NoStop}%
\bibitem [{\citenamefont {Karamitsos}\ and\ \citenamefont
  {Pilaftsis}(2018)}]{Karamitsos2018}%
  \BibitemOpen
  \bibfield  {author} {\bibinfo {author} {\bibfnamefont {Sotirios}\
  \bibnamefont {Karamitsos}}\ and\ \bibinfo {author} {\bibfnamefont
  {Apostolos}\ \bibnamefont {Pilaftsis}},\ }\bibfield  {title} {\enquote
  {\bibinfo {title} {{On the Cosmological Frame Problem}},}\ }\bibfield
  {booktitle} {\emph {\bibinfo {booktitle} {{Proceedings, 17th Hellenic School
  and Workshops on Elementary Particle Physics and Gravity (CORFU2017): Corfu,
  Greece, September 2-28, 2017}}},\ }\href {\doibase 10.22323/1.318.0036}
  {\bibfield  {journal} {\bibinfo  {journal} {PoS}\ }\textbf {\bibinfo {volume}
  {CORFU2017}},\ \bibinfo {pages} {036} (\bibinfo {year} {2018})},\ \Eprint
  {http://arxiv.org/abs/1801.07151} {arXiv:1801.07151 [hep-th]} \BibitemShut
  {NoStop}%
\bibitem [{\citenamefont {Bhattacharya}\ \emph {et~al.}(2018)\citenamefont
  {Bhattacharya}, \citenamefont {Das},\ and\ \citenamefont
  {Majhi}}]{Bhattacharya2018}%
  \BibitemOpen
  \bibfield  {author} {\bibinfo {author} {\bibfnamefont {Krishnakanta}\
  \bibnamefont {Bhattacharya}}, \bibinfo {author} {\bibfnamefont {Ashmita}\
  \bibnamefont {Das}}, \ and\ \bibinfo {author} {\bibfnamefont {Bibhas~Ranjan}\
  \bibnamefont {Majhi}},\ }\bibfield  {title} {\enquote {\bibinfo {title}
  {{Noether and Abbott-Deser-Tekin conserved quantities in scalar-tensor theory
  of gravity both in Jordan and Einstein frames}},}\ }\href {\doibase
  10.1103/PhysRevD.97.124013} {\bibfield  {journal} {\bibinfo  {journal} {Phys.
  Rev.}\ }\textbf {\bibinfo {volume} {D97}},\ \bibinfo {pages} {124013}
  (\bibinfo {year} {2018})},\ \Eprint {http://arxiv.org/abs/1803.03771}
  {arXiv:1803.03771 [gr-qc]} \BibitemShut {NoStop}%
\bibitem [{\citenamefont {Quiros}\ and\ \citenamefont
  {De~Arcia}(2018)}]{Quiros2018}%
  \BibitemOpen
  \bibfield  {author} {\bibinfo {author} {\bibfnamefont {Israel}\ \bibnamefont
  {Quiros}}\ and\ \bibinfo {author} {\bibfnamefont {Roberto}\ \bibnamefont
  {De~Arcia}},\ }\bibfield  {title} {\enquote {\bibinfo {title} {{On local
  scale invariance and the questionable theoretical basis of the conformal
  transformations' issue}},}\ }\href@noop {} {\  (\bibinfo {year} {2018})},\
  \Eprint {http://arxiv.org/abs/1811.02458} {arXiv:1811.02458 [gr-qc]}
  \BibitemShut {NoStop}%
\bibitem [{\citenamefont {Falls}\ and\ \citenamefont
  {Herrero-Valea}(2018)}]{Falls2018}%
  \BibitemOpen
  \bibfield  {author} {\bibinfo {author} {\bibfnamefont {Kevin}\ \bibnamefont
  {Falls}}\ and\ \bibinfo {author} {\bibfnamefont {Mario}\ \bibnamefont
  {Herrero-Valea}},\ }\bibfield  {title} {\enquote {\bibinfo {title} {{Frame
  (In)equivalence in Quantum Field Theory and Cosmology}},}\ }\href@noop {} {\
  (\bibinfo {year} {2018})},\ \Eprint {http://arxiv.org/abs/1812.08187}
  {arXiv:1812.08187 [hep-th]} \BibitemShut {NoStop}%
\bibitem [{\citenamefont {Chakraborty}\ \emph {et~al.}(2019)\citenamefont
  {Chakraborty}, \citenamefont {Pal},\ and\ \citenamefont
  {Saa}}]{Chakraborty2019}%
  \BibitemOpen
  \bibfield  {author} {\bibinfo {author} {\bibfnamefont {Saikat}\ \bibnamefont
  {Chakraborty}}, \bibinfo {author} {\bibfnamefont {Sanchari}\ \bibnamefont
  {Pal}}, \ and\ \bibinfo {author} {\bibfnamefont {Alberto}\ \bibnamefont
  {Saa}},\ }\bibfield  {title} {\enquote {\bibinfo {title} {{Dynamical
  equivalence of $f(R)$ gravity in Jordan and Einstein frames}},}\ }\href
  {\doibase 10.1103/PhysRevD.99.024020} {\bibfield  {journal} {\bibinfo
  {journal} {Phys. Rev.}\ }\textbf {\bibinfo {volume} {D99}},\ \bibinfo {pages}
  {024020} (\bibinfo {year} {2019})},\ \Eprint
  {http://arxiv.org/abs/1812.01694} {arXiv:1812.01694 [gr-qc]} \BibitemShut
  {NoStop}%
\bibitem [{\citenamefont {Quiros}(2019)}]{Quiros2019}%
  \BibitemOpen
  \bibfield  {author} {\bibinfo {author} {\bibfnamefont {Israel}\ \bibnamefont
  {Quiros}},\ }\bibfield  {title} {\enquote {\bibinfo {title} {{Selected topics
  in scalar-tensor theories and beyond}},}\ }\href {\doibase
  10.1142/S021827181930012X} {\  (\bibinfo {year} {2019}),\
  10.1142/S021827181930012X},\ \Eprint {http://arxiv.org/abs/1901.08690}
  {arXiv:1901.08690 [gr-qc]} \BibitemShut {NoStop}%
\bibitem [{\citenamefont {Karwan}\ and\ \citenamefont
  {Channuie}(2019)}]{Karwan:2018eln}%
  \BibitemOpen
  \bibfield  {author} {\bibinfo {author} {\bibfnamefont {Khamphee}\
  \bibnamefont {Karwan}}\ and\ \bibinfo {author} {\bibfnamefont {Phongpichit}\
  \bibnamefont {Channuie}},\ }\bibfield  {title} {\enquote {\bibinfo {title}
  {{Generalized Conformal Transformation and Inflationary Attractors}},}\
  }\href {\doibase 10.1103/PhysRevD.100.023514} {\bibfield  {journal} {\bibinfo
   {journal} {Phys. Rev. D}\ }\textbf {\bibinfo {volume} {100}},\ \bibinfo
  {pages} {023514} (\bibinfo {year} {2019})},\ \Eprint
  {http://arxiv.org/abs/1811.03006} {arXiv:1811.03006 [gr-qc]} \BibitemShut
  {NoStop}%
\bibitem [{\citenamefont {Nandi}(2019)}]{Nandi:2019xlj}%
  \BibitemOpen
  \bibfield  {author} {\bibinfo {author} {\bibfnamefont {Debottam}\
  \bibnamefont {Nandi}},\ }\bibfield  {title} {\enquote {\bibinfo {title}
  {{Note on stability in conformally connected frames}},}\ }\href {\doibase
  10.1103/PhysRevD.99.103532} {\bibfield  {journal} {\bibinfo  {journal} {Phys.
  Rev. D}\ }\textbf {\bibinfo {volume} {99}},\ \bibinfo {pages} {103532}
  (\bibinfo {year} {2019})},\ \Eprint {http://arxiv.org/abs/1904.00153}
  {arXiv:1904.00153 [gr-qc]} \BibitemShut {NoStop}%
\bibitem [{\citenamefont {Ema}(2019)}]{Ema:2019fdd}%
  \BibitemOpen
  \bibfield  {author} {\bibinfo {author} {\bibfnamefont {Yohei}\ \bibnamefont
  {Ema}},\ }\bibfield  {title} {\enquote {\bibinfo {title} {{Dynamical
  Emergence of Scalaron in Higgs Inflation}},}\ }\href {\doibase
  10.1088/1475-7516/2019/09/027} {\bibfield  {journal} {\bibinfo  {journal}
  {JCAP}\ }\textbf {\bibinfo {volume} {09}},\ \bibinfo {pages} {027} (\bibinfo
  {year} {2019})},\ \Eprint {http://arxiv.org/abs/1907.00993} {arXiv:1907.00993
  [hep-ph]} \BibitemShut {NoStop}%
\bibitem [{\citenamefont {Nashed}\ \emph {et~al.}(2019)\citenamefont {Nashed},
  \citenamefont {El~Hanafy}, \citenamefont {Odintsov},\ and\ \citenamefont
  {Oikonomou}}]{Nashed:2019yto}%
  \BibitemOpen
  \bibfield  {author} {\bibinfo {author} {\bibfnamefont {G.G.L.}\ \bibnamefont
  {Nashed}}, \bibinfo {author} {\bibfnamefont {W.}~\bibnamefont {El~Hanafy}},
  \bibinfo {author} {\bibfnamefont {S.D.}\ \bibnamefont {Odintsov}}, \ and\
  \bibinfo {author} {\bibfnamefont {V.K.}\ \bibnamefont {Oikonomou}},\
  }\bibfield  {title} {\enquote {\bibinfo {title} {{Thermodynamical
  correspondence of $f(R)$ gravity in Jordan and Einstein frames}},}\
  }\href@noop {} {\  (\bibinfo {year} {2019})},\ \Eprint
  {http://arxiv.org/abs/1912.03897} {arXiv:1912.03897 [gr-qc]} \BibitemShut
  {NoStop}%
\bibitem [{\citenamefont {J\"{a}rv}\ \emph {et~al.}(2015)\citenamefont
  {J\"{a}rv}, \citenamefont {Kuusk}, \citenamefont {Saal},\ and\ \citenamefont
  {Vilson}}]{Jaerv2015}%
  \BibitemOpen
  \bibfield  {author} {\bibinfo {author} {\bibfnamefont {Laur}\ \bibnamefont
  {J\"{a}rv}}, \bibinfo {author} {\bibfnamefont {Piret}\ \bibnamefont {Kuusk}},
  \bibinfo {author} {\bibfnamefont {Margus}\ \bibnamefont {Saal}}, \ and\
  \bibinfo {author} {\bibfnamefont {Ott}\ \bibnamefont {Vilson}},\ }\bibfield
  {title} {\enquote {\bibinfo {title} {Invariant quantities in the
  scalar-tensor theories of gravitation},}\ }\href {\doibase
  10.1103/PhysRevD.91.024041} {\bibfield  {journal} {\bibinfo  {journal}
  {Physical Review D}\ }\textbf {\bibinfo {volume} {91}},\ \bibinfo {pages}
  {024041} (\bibinfo {year} {2015})},\ \Eprint {http://arxiv.org/abs/1411.1947}
  {arXiv:1411.1947 [gr-qc]} \BibitemShut {NoStop}%
\bibitem [{\citenamefont {J{\"a}rv}\ \emph {et~al.}(2015)\citenamefont
  {J{\"a}rv}, \citenamefont {Kuusk}, \citenamefont {Saal},\ and\ \citenamefont
  {Vilson}}]{Jaerv2015a}%
  \BibitemOpen
  \bibfield  {author} {\bibinfo {author} {\bibfnamefont {Laur}\ \bibnamefont
  {J{\"a}rv}}, \bibinfo {author} {\bibfnamefont {Piret}\ \bibnamefont {Kuusk}},
  \bibinfo {author} {\bibfnamefont {Margus}\ \bibnamefont {Saal}}, \ and\
  \bibinfo {author} {\bibfnamefont {Ott}\ \bibnamefont {Vilson}},\ }\bibfield
  {title} {\enquote {\bibinfo {title} {Transformation properties and general
  relativity regime in scalar--tensor theories},}\ }\href {\doibase
  10.1088/0264-9381/32/23/235013} {\bibfield  {journal} {\bibinfo  {journal}
  {Classical and Quantum Gravity}\ }\textbf {\bibinfo {volume} {32}},\ \bibinfo
  {pages} {235013} (\bibinfo {year} {2015})},\ \Eprint
  {http://arxiv.org/abs/1504.02686} {arXiv:1504.02686 [gr-qc]} \BibitemShut
  {NoStop}%
\bibitem [{\citenamefont {Jarv}\ \emph {et~al.}(2015)\citenamefont {Jarv},
  \citenamefont {Kuusk}, \citenamefont {Saal},\ and\ \citenamefont
  {Vilson}}]{Jarv2015}%
  \BibitemOpen
  \bibfield  {author} {\bibinfo {author} {\bibfnamefont {Laur}\ \bibnamefont
  {Jarv}}, \bibinfo {author} {\bibfnamefont {Piret}\ \bibnamefont {Kuusk}},
  \bibinfo {author} {\bibfnamefont {Margus}\ \bibnamefont {Saal}}, \ and\
  \bibinfo {author} {\bibfnamefont {Ott}\ \bibnamefont {Vilson}},\ }\bibfield
  {title} {\enquote {\bibinfo {title} {{The formalism of invariants in
  scalar-tensor and multiscalar-tensor theories of gravitation}},}\ }in\ \href
  {http://inspirehep.net/record/1411793/files/arXiv:1512.09166.pdf} {\emph
  {\bibinfo {booktitle} {{14th Marcel Grossmann Meeting on Recent Developments
  in Theoretical and Experimental General Relativity, Astrophysics, and
  Relativistic Field Theories (MG14) Rome, Italy, July 12-18, 2015}}}}\
  (\bibinfo {year} {2015})\ \Eprint {http://arxiv.org/abs/1512.09166}
  {arXiv:1512.09166 [gr-qc]} \BibitemShut {NoStop}%
\bibitem [{\citenamefont {Kuusk}\ \emph
  {et~al.}(2016{\natexlab{a}})\citenamefont {Kuusk}, \citenamefont
  {R{\"u}nkla}, \citenamefont {Saal},\ and\ \citenamefont
  {Vilson}}]{Kuusk2016}%
  \BibitemOpen
  \bibfield  {author} {\bibinfo {author} {\bibfnamefont {Piret}\ \bibnamefont
  {Kuusk}}, \bibinfo {author} {\bibfnamefont {Mihkel}\ \bibnamefont
  {R{\"u}nkla}}, \bibinfo {author} {\bibfnamefont {Margus}\ \bibnamefont
  {Saal}}, \ and\ \bibinfo {author} {\bibfnamefont {Ott}\ \bibnamefont
  {Vilson}},\ }\bibfield  {title} {\enquote {\bibinfo {title} {{Invariant
  slow-roll parameters in scalar–tensor theories}},}\ }\href {\doibase
  10.1088/0264-9381/33/19/195008} {\bibfield  {journal} {\bibinfo  {journal}
  {Class. Quant. Grav.}\ }\textbf {\bibinfo {volume} {33}},\ \bibinfo {pages}
  {195008} (\bibinfo {year} {2016}{\natexlab{a}})},\ \Eprint
  {http://arxiv.org/abs/1605.07033} {arXiv:1605.07033 [gr-qc]} \BibitemShut
  {NoStop}%
\bibitem [{\citenamefont {J{\"a}rv}\ \emph {et~al.}(2017)\citenamefont
  {J{\"a}rv}, \citenamefont {Kannike}, \citenamefont {Marzola}, \citenamefont
  {Racioppi}, \citenamefont {Raidal}, \citenamefont {R{\"u}nkla}, \citenamefont
  {Saal},\ and\ \citenamefont {Veerm{\"a}e}}]{Jarv2017}%
  \BibitemOpen
  \bibfield  {author} {\bibinfo {author} {\bibfnamefont {Laur}\ \bibnamefont
  {J{\"a}rv}}, \bibinfo {author} {\bibfnamefont {Kristjan}\ \bibnamefont
  {Kannike}}, \bibinfo {author} {\bibfnamefont {Luca}\ \bibnamefont {Marzola}},
  \bibinfo {author} {\bibfnamefont {Antonio}\ \bibnamefont {Racioppi}},
  \bibinfo {author} {\bibfnamefont {Martti}\ \bibnamefont {Raidal}}, \bibinfo
  {author} {\bibfnamefont {Mihkel}\ \bibnamefont {R{\"u}nkla}}, \bibinfo
  {author} {\bibfnamefont {Margus}\ \bibnamefont {Saal}}, \ and\ \bibinfo
  {author} {\bibfnamefont {Hardi}\ \bibnamefont {Veerm{\"a}e}},\ }\bibfield
  {title} {\enquote {\bibinfo {title} {{Frame-Independent Classification of
  Single-Field Inflationary Models}},}\ }\href {\doibase
  10.1103/PhysRevLett.118.151302} {\bibfield  {journal} {\bibinfo  {journal}
  {Phys. Rev. Lett.}\ }\textbf {\bibinfo {volume} {118}},\ \bibinfo {pages}
  {151302} (\bibinfo {year} {2017})},\ \Eprint
  {http://arxiv.org/abs/1612.06863} {arXiv:1612.06863 [hep-ph]} \BibitemShut
  {NoStop}%
\bibitem [{\citenamefont {Karam}\ \emph {et~al.}(2017)\citenamefont {Karam},
  \citenamefont {Pappas},\ and\ \citenamefont {Tamvakis}}]{Karam2017}%
  \BibitemOpen
  \bibfield  {author} {\bibinfo {author} {\bibfnamefont {Alexandros}\
  \bibnamefont {Karam}}, \bibinfo {author} {\bibfnamefont {Thomas}\
  \bibnamefont {Pappas}}, \ and\ \bibinfo {author} {\bibfnamefont {Kyriakos}\
  \bibnamefont {Tamvakis}},\ }\bibfield  {title} {\enquote {\bibinfo {title}
  {{Frame-dependence of higher-order inflationary observables in scalar-tensor
  theories}},}\ }\href {\doibase 10.1103/PhysRevD.96.064036} {\bibfield
  {journal} {\bibinfo  {journal} {Phys. Rev.}\ }\textbf {\bibinfo {volume}
  {D96}},\ \bibinfo {pages} {064036} (\bibinfo {year} {2017})},\ \Eprint
  {http://arxiv.org/abs/1707.00984} {arXiv:1707.00984 [gr-qc]} \BibitemShut
  {NoStop}%
\bibitem [{\citenamefont {Kuusk}\ \emph
  {et~al.}(2016{\natexlab{b}})\citenamefont {Kuusk}, \citenamefont {J\"{a}rv},\
  and\ \citenamefont {Vilson}}]{Kuusk2016a}%
  \BibitemOpen
  \bibfield  {author} {\bibinfo {author} {\bibfnamefont {Piret}\ \bibnamefont
  {Kuusk}}, \bibinfo {author} {\bibfnamefont {Laur}\ \bibnamefont {J\"{a}rv}},
  \ and\ \bibinfo {author} {\bibfnamefont {Ott}\ \bibnamefont {Vilson}},\
  }\bibfield  {title} {\enquote {\bibinfo {title} {{Invariant quantities in the
  multiscalar-tensor theories of gravitation}},}\ }\href {\doibase
  10.1142/S0217751X16410037} {\bibfield  {journal} {\bibinfo  {journal} {Int.
  J. Mod. Phys.}\ }\textbf {\bibinfo {volume} {A31}},\ \bibinfo {pages}
  {1641003} (\bibinfo {year} {2016}{\natexlab{b}})},\ \Eprint
  {http://arxiv.org/abs/1509.02903} {arXiv:1509.02903 [gr-qc]} \BibitemShut
  {NoStop}%
\bibitem [{\citenamefont {Karam}\ \emph {et~al.}(2018)\citenamefont {Karam},
  \citenamefont {Lykkas},\ and\ \citenamefont {Tamvakis}}]{Karam2018}%
  \BibitemOpen
  \bibfield  {author} {\bibinfo {author} {\bibfnamefont {Alexandros}\
  \bibnamefont {Karam}}, \bibinfo {author} {\bibfnamefont {Angelos}\
  \bibnamefont {Lykkas}}, \ and\ \bibinfo {author} {\bibfnamefont {Kyriakos}\
  \bibnamefont {Tamvakis}},\ }\bibfield  {title} {\enquote {\bibinfo {title}
  {{Frame-invariant approach to higher-dimensional scalar-tensor gravity}},}\
  }\href {\doibase 10.1103/PhysRevD.97.124036} {\bibfield  {journal} {\bibinfo
  {journal} {Phys. Rev.}\ }\textbf {\bibinfo {volume} {D97}},\ \bibinfo {pages}
  {124036} (\bibinfo {year} {2018})},\ \Eprint
  {http://arxiv.org/abs/1803.04960} {arXiv:1803.04960 [gr-qc]} \BibitemShut
  {NoStop}%
\bibitem [{\citenamefont {Kozak}\ and\ \citenamefont
  {Borowiec}(2019)}]{Kozak:2018vlp}%
  \BibitemOpen
  \bibfield  {author} {\bibinfo {author} {\bibfnamefont {Aleksander}\
  \bibnamefont {Kozak}}\ and\ \bibinfo {author} {\bibfnamefont {Andrzej}\
  \bibnamefont {Borowiec}},\ }\bibfield  {title} {\enquote {\bibinfo {title}
  {{Palatini frames in scalar–tensor theories of gravity}},}\ }\href
  {\doibase 10.1140/epjc/s10052-019-6836-y} {\bibfield  {journal} {\bibinfo
  {journal} {Eur. Phys. J.}\ }\textbf {\bibinfo {volume} {C79}},\ \bibinfo
  {pages} {335} (\bibinfo {year} {2019})},\ \Eprint
  {http://arxiv.org/abs/1808.05598} {arXiv:1808.05598 [hep-th]} \BibitemShut
  {NoStop}%
\bibitem [{\citenamefont {Borowiec}\ and\ \citenamefont
  {Kozak}(2020)}]{Borowiec:2020lfx}%
  \BibitemOpen
  \bibfield  {author} {\bibinfo {author} {\bibfnamefont {Andrzej}\ \bibnamefont
  {Borowiec}}\ and\ \bibinfo {author} {\bibfnamefont {Aleksander}\ \bibnamefont
  {Kozak}},\ }\bibfield  {title} {\enquote {\bibinfo {title} {{New class of
  hybrid metric-Palatini scalar-tensor theories of gravity}},}\ }\href@noop {}
  {\  (\bibinfo {year} {2020})},\ \Eprint {http://arxiv.org/abs/2003.02741}
  {arXiv:2003.02741 [gr-qc]} \BibitemShut {NoStop}%
\bibitem [{\citenamefont {Hohmann}(2018)}]{Hohmann2018}%
  \BibitemOpen
  \bibfield  {author} {\bibinfo {author} {\bibfnamefont {Manuel}\ \bibnamefont
  {Hohmann}},\ }\bibfield  {title} {\enquote {\bibinfo {title} {{Scalar-torsion
  theories of gravity III: analogue of scalar-tensor gravity and conformal
  invariants}},}\ }\href {\doibase 10.1103/PhysRevD.98.064004} {\bibfield
  {journal} {\bibinfo  {journal} {Phys. Rev.}\ }\textbf {\bibinfo {volume}
  {D98}},\ \bibinfo {pages} {064004} (\bibinfo {year} {2018})},\ \Eprint
  {http://arxiv.org/abs/1801.06531} {arXiv:1801.06531 [gr-qc]} \BibitemShut
  {NoStop}%
\bibitem [{\citenamefont {Palatini}(1919)}]{Palatini1919}%
  \BibitemOpen
  \bibfield  {author} {\bibinfo {author} {\bibfnamefont {Attilio}\ \bibnamefont
  {Palatini}},\ }\bibfield  {title} {\enquote {\bibinfo {title} {Deduzione
  invariantiva delle equazioni gravitazionali dal principio di hamilton},}\
  }\href {\doibase 10.1007/BF03014670} {\bibfield  {journal} {\bibinfo
  {journal} {Rendiconti del Circolo Matematico di Palermo (1884-1940)}\
  }\textbf {\bibinfo {volume} {43}},\ \bibinfo {pages} {203--212} (\bibinfo
  {year} {1919})}\BibitemShut {NoStop}%
\bibitem [{\citenamefont {Ferraris}\ \emph {et~al.}(1982)\citenamefont
  {Ferraris}, \citenamefont {Francaviglia},\ and\ \citenamefont
  {Reina}}]{Ferraris1982}%
  \BibitemOpen
  \bibfield  {author} {\bibinfo {author} {\bibfnamefont {M.}~\bibnamefont
  {Ferraris}}, \bibinfo {author} {\bibfnamefont {M.}~\bibnamefont
  {Francaviglia}}, \ and\ \bibinfo {author} {\bibfnamefont {C.}~\bibnamefont
  {Reina}},\ }\bibfield  {title} {\enquote {\bibinfo {title} {Variational
  formulation of general relativity from 1915 to 1925 ``palatini's method''
  discovered by einstein in 1925},}\ }\href {\doibase 10.1007/BF00756060}
  {\bibfield  {journal} {\bibinfo  {journal} {General Relativity and
  Gravitation}\ }\textbf {\bibinfo {volume} {14}},\ \bibinfo {pages} {243--254}
  (\bibinfo {year} {1982})}\BibitemShut {NoStop}%
\bibitem [{\citenamefont {Exirifard}\ and\ \citenamefont
  {Sheikh-Jabbari}(2008)}]{Exirifard2008}%
  \BibitemOpen
  \bibfield  {author} {\bibinfo {author} {\bibfnamefont {Q.}~\bibnamefont
  {Exirifard}}\ and\ \bibinfo {author} {\bibfnamefont {M.~M.}\ \bibnamefont
  {Sheikh-Jabbari}},\ }\bibfield  {title} {\enquote {\bibinfo {title}
  {{Lovelock gravity at the crossroads of Palatini and metric formulations}},}\
  }\href {\doibase 10.1016/j.physletb.2008.02.012} {\bibfield  {journal}
  {\bibinfo  {journal} {Phys. Lett.}\ }\textbf {\bibinfo {volume} {B661}},\
  \bibinfo {pages} {158--161} (\bibinfo {year} {2008})},\ \Eprint
  {http://arxiv.org/abs/0705.1879} {arXiv:0705.1879 [hep-th]} \BibitemShut
  {NoStop}%
\bibitem [{\citenamefont {Bauer}\ and\ \citenamefont
  {Demir}(2008)}]{Bauer2008}%
  \BibitemOpen
  \bibfield  {author} {\bibinfo {author} {\bibfnamefont {Florian}\ \bibnamefont
  {Bauer}}\ and\ \bibinfo {author} {\bibfnamefont {Durmus~A.}\ \bibnamefont
  {Demir}},\ }\bibfield  {title} {\enquote {\bibinfo {title} {{Inflation with
  Non-Minimal Coupling: Metric versus Palatini Formulations}},}\ }\href
  {\doibase 10.1016/j.physletb.2008.06.014} {\bibfield  {journal} {\bibinfo
  {journal} {Phys. Lett.}\ }\textbf {\bibinfo {volume} {B665}},\ \bibinfo
  {pages} {222--226} (\bibinfo {year} {2008})},\ \Eprint
  {http://arxiv.org/abs/0803.2664} {arXiv:0803.2664 [hep-ph]} \BibitemShut
  {NoStop}%
\bibitem [{\citenamefont {Bauer}(2011)}]{Bauer2011}%
  \BibitemOpen
  \bibfield  {author} {\bibinfo {author} {\bibfnamefont {Florian}\ \bibnamefont
  {Bauer}},\ }\bibfield  {title} {\enquote {\bibinfo {title} {{Filtering out
  the cosmological constant in the Palatini formalism of modified gravity}},}\
  }\href {\doibase 10.1007/s10714-011-1153-2} {\bibfield  {journal} {\bibinfo
  {journal} {Gen. Rel. Grav.}\ }\textbf {\bibinfo {volume} {43}},\ \bibinfo
  {pages} {1733--1757} (\bibinfo {year} {2011})},\ \Eprint
  {http://arxiv.org/abs/1007.2546} {arXiv:1007.2546 [gr-qc]} \BibitemShut
  {NoStop}%
\bibitem [{\citenamefont {Tamanini}\ and\ \citenamefont
  {Contaldi}(2011)}]{Tamanini2011}%
  \BibitemOpen
  \bibfield  {author} {\bibinfo {author} {\bibfnamefont {Nicola}\ \bibnamefont
  {Tamanini}}\ and\ \bibinfo {author} {\bibfnamefont {Carlo~R.}\ \bibnamefont
  {Contaldi}},\ }\bibfield  {title} {\enquote {\bibinfo {title} {{Inflationary
  Perturbations in Palatini Generalised Gravity}},}\ }\href {\doibase
  10.1103/PhysRevD.83.044018} {\bibfield  {journal} {\bibinfo  {journal} {Phys.
  Rev.}\ }\textbf {\bibinfo {volume} {D83}},\ \bibinfo {pages} {044018}
  (\bibinfo {year} {2011})},\ \Eprint {http://arxiv.org/abs/1010.0689}
  {arXiv:1010.0689 [gr-qc]} \BibitemShut {NoStop}%
\bibitem [{\citenamefont {Bauer}\ and\ \citenamefont
  {Demir}(2011)}]{Bauer:2010jg}%
  \BibitemOpen
  \bibfield  {author} {\bibinfo {author} {\bibfnamefont {Florian}\ \bibnamefont
  {Bauer}}\ and\ \bibinfo {author} {\bibfnamefont {Durmus~A.}\ \bibnamefont
  {Demir}},\ }\bibfield  {title} {\enquote {\bibinfo {title} {{Higgs-Palatini
  Inflation and Unitarity}},}\ }\href {\doibase 10.1016/j.physletb.2011.03.042}
  {\bibfield  {journal} {\bibinfo  {journal} {Phys. Lett.}\ }\textbf {\bibinfo
  {volume} {B698}},\ \bibinfo {pages} {425--429} (\bibinfo {year} {2011})},\
  \Eprint {http://arxiv.org/abs/1012.2900} {arXiv:1012.2900 [hep-ph]}
  \BibitemShut {NoStop}%
\bibitem [{\citenamefont {Olmo}(2011)}]{Olmo2011}%
  \BibitemOpen
  \bibfield  {author} {\bibinfo {author} {\bibfnamefont {Gonzalo~J.}\
  \bibnamefont {Olmo}},\ }\bibfield  {title} {\enquote {\bibinfo {title}
  {{Palatini Approach to Modified Gravity: f(R) Theories and Beyond}},}\ }\href
  {\doibase 10.1142/S0218271811018925} {\bibfield  {journal} {\bibinfo
  {journal} {Int. J. Mod. Phys.}\ }\textbf {\bibinfo {volume} {D20}},\ \bibinfo
  {pages} {413--462} (\bibinfo {year} {2011})},\ \Eprint
  {http://arxiv.org/abs/1101.3864} {arXiv:1101.3864 [gr-qc]} \BibitemShut
  {NoStop}%
\bibitem [{\citenamefont {Rasanen}\ and\ \citenamefont
  {Wahlman}(2017)}]{Rasanen2017}%
  \BibitemOpen
  \bibfield  {author} {\bibinfo {author} {\bibfnamefont {Syksy}\ \bibnamefont
  {Rasanen}}\ and\ \bibinfo {author} {\bibfnamefont {Pyry}\ \bibnamefont
  {Wahlman}},\ }\bibfield  {title} {\enquote {\bibinfo {title} {{Higgs
  inflation with loop corrections in the Palatini formulation}},}\ }\href
  {\doibase 10.1088/1475-7516/2017/11/047} {\bibfield  {journal} {\bibinfo
  {journal} {JCAP}\ }\textbf {\bibinfo {volume} {11}},\ \bibinfo {pages} {047}
  (\bibinfo {year} {2017})},\ \Eprint {http://arxiv.org/abs/1709.07853}
  {arXiv:1709.07853 [astro-ph.CO]} \BibitemShut {NoStop}%
\bibitem [{\citenamefont {Racioppi}(2017)}]{Racioppi2017}%
  \BibitemOpen
  \bibfield  {author} {\bibinfo {author} {\bibfnamefont {Antonio}\ \bibnamefont
  {Racioppi}},\ }\bibfield  {title} {\enquote {\bibinfo {title}
  {{Coleman-Weinberg linear inflation: metric vs. Palatini formulation}},}\
  }\href {\doibase 10.1088/1475-7516/2017/12/041} {\bibfield  {journal}
  {\bibinfo  {journal} {JCAP}\ }\textbf {\bibinfo {volume} {1712}},\ \bibinfo
  {pages} {041} (\bibinfo {year} {2017})},\ \Eprint
  {http://arxiv.org/abs/1710.04853} {arXiv:1710.04853 [astro-ph.CO]}
  \BibitemShut {NoStop}%
\bibitem [{\citenamefont {Racioppi}(2018)}]{Racioppi2018}%
  \BibitemOpen
  \bibfield  {author} {\bibinfo {author} {\bibfnamefont {Antonio}\ \bibnamefont
  {Racioppi}},\ }\bibfield  {title} {\enquote {\bibinfo {title} {{New universal
  attractor in nonminimally coupled gravity: Linear inflation}},}\ }\href
  {\doibase 10.1103/PhysRevD.97.123514} {\bibfield  {journal} {\bibinfo
  {journal} {Phys. Rev.}\ }\textbf {\bibinfo {volume} {D97}},\ \bibinfo {pages}
  {123514} (\bibinfo {year} {2018})},\ \Eprint
  {http://arxiv.org/abs/1801.08810} {arXiv:1801.08810 [astro-ph.CO]}
  \BibitemShut {NoStop}%
\bibitem [{\citenamefont {Järv}\ \emph {et~al.}(2018)\citenamefont {Järv},
  \citenamefont {Racioppi},\ and\ \citenamefont {Tenkanen}}]{Jaerv2018}%
  \BibitemOpen
  \bibfield  {author} {\bibinfo {author} {\bibfnamefont {Laur}\ \bibnamefont
  {Järv}}, \bibinfo {author} {\bibfnamefont {Antonio}\ \bibnamefont
  {Racioppi}}, \ and\ \bibinfo {author} {\bibfnamefont {Tommi}\ \bibnamefont
  {Tenkanen}},\ }\bibfield  {title} {\enquote {\bibinfo {title} {{Palatini side
  of inflationary attractors}},}\ }\href {\doibase 10.1103/PhysRevD.97.083513}
  {\bibfield  {journal} {\bibinfo  {journal} {Phys. Rev. D}\ }\textbf {\bibinfo
  {volume} {97}},\ \bibinfo {pages} {083513} (\bibinfo {year} {2018})},\
  \Eprint {http://arxiv.org/abs/1712.08471} {arXiv:1712.08471 [gr-qc]}
  \BibitemShut {NoStop}%
\bibitem [{\citenamefont {Bombacigno}\ and\ \citenamefont
  {Montani}(2019)}]{Bombacigno2019}%
  \BibitemOpen
  \bibfield  {author} {\bibinfo {author} {\bibfnamefont {Flavio}\ \bibnamefont
  {Bombacigno}}\ and\ \bibinfo {author} {\bibfnamefont {Giovanni}\ \bibnamefont
  {Montani}},\ }\bibfield  {title} {\enquote {\bibinfo {title} {{Big bounce
  cosmology for Palatini $R^2$ gravity with a Nieh--Yan term}},}\ }\href
  {\doibase 10.1140/epjc/s10052-019-6918-x} {\bibfield  {journal} {\bibinfo
  {journal} {Eur. Phys. J. C}\ }\textbf {\bibinfo {volume} {79}},\ \bibinfo
  {pages} {405} (\bibinfo {year} {2019})},\ \Eprint
  {http://arxiv.org/abs/1809.07563} {arXiv:1809.07563 [gr-qc]} \BibitemShut
  {NoStop}%
\bibitem [{\citenamefont {Rasanen}\ and\ \citenamefont
  {Tomberg}(2019)}]{Rasanen2019}%
  \BibitemOpen
  \bibfield  {author} {\bibinfo {author} {\bibfnamefont {Syksy}\ \bibnamefont
  {Rasanen}}\ and\ \bibinfo {author} {\bibfnamefont {Eemeli}\ \bibnamefont
  {Tomberg}},\ }\bibfield  {title} {\enquote {\bibinfo {title} {{Planck scale
  black hole dark matter from Higgs inflation}},}\ }\href {\doibase
  10.1088/1475-7516/2019/01/038} {\bibfield  {journal} {\bibinfo  {journal}
  {JCAP}\ }\textbf {\bibinfo {volume} {01}},\ \bibinfo {pages} {038} (\bibinfo
  {year} {2019})},\ \Eprint {http://arxiv.org/abs/1810.12608} {arXiv:1810.12608
  [astro-ph.CO]} \BibitemShut {NoStop}%
\bibitem [{\citenamefont {Rasanen}(2018)}]{Rasanen2018}%
  \BibitemOpen
  \bibfield  {author} {\bibinfo {author} {\bibfnamefont {Syksy}\ \bibnamefont
  {Rasanen}},\ }\bibfield  {title} {\enquote {\bibinfo {title} {{Higgs
  inflation in the Palatini formulation with kinetic terms for the metric}},}\
  }\href {\doibase 10.21105/astro.1811.09514} {\  (\bibinfo {year} {2018}),\
  10.21105/astro.1811.09514},\ \Eprint {http://arxiv.org/abs/1811.09514}
  {arXiv:1811.09514 [gr-qc]} \BibitemShut {NoStop}%
\bibitem [{\citenamefont {Almeida}\ \emph {et~al.}(2019)\citenamefont
  {Almeida}, \citenamefont {Bernal}, \citenamefont {Rubio},\ and\ \citenamefont
  {Tenkanen}}]{Almeida2019}%
  \BibitemOpen
  \bibfield  {author} {\bibinfo {author} {\bibfnamefont {Juan P.~Beltrán}\
  \bibnamefont {Almeida}}, \bibinfo {author} {\bibfnamefont {Nicolás}\
  \bibnamefont {Bernal}}, \bibinfo {author} {\bibfnamefont {Javier}\
  \bibnamefont {Rubio}}, \ and\ \bibinfo {author} {\bibfnamefont {Tommi}\
  \bibnamefont {Tenkanen}},\ }\bibfield  {title} {\enquote {\bibinfo {title}
  {{Hidden Inflaton Dark Matter}},}\ }\href {\doibase
  10.1088/1475-7516/2019/03/012} {\bibfield  {journal} {\bibinfo  {journal}
  {JCAP}\ }\textbf {\bibinfo {volume} {03}},\ \bibinfo {pages} {012} (\bibinfo
  {year} {2019})},\ \Eprint {http://arxiv.org/abs/1811.09640} {arXiv:1811.09640
  [hep-ph]} \BibitemShut {NoStop}%
\bibitem [{\citenamefont {Shimada}\ \emph {et~al.}(2019)\citenamefont
  {Shimada}, \citenamefont {Aoki},\ and\ \citenamefont {Maeda}}]{Shimada2019}%
  \BibitemOpen
  \bibfield  {author} {\bibinfo {author} {\bibfnamefont {Keigo}\ \bibnamefont
  {Shimada}}, \bibinfo {author} {\bibfnamefont {Katsuki}\ \bibnamefont {Aoki}},
  \ and\ \bibinfo {author} {\bibfnamefont {Kei-ichi}\ \bibnamefont {Maeda}},\
  }\bibfield  {title} {\enquote {\bibinfo {title} {{Metric-affine Gravity and
  Inflation}},}\ }\href {\doibase 10.1103/PhysRevD.99.104020} {\bibfield
  {journal} {\bibinfo  {journal} {Phys. Rev. D}\ }\textbf {\bibinfo {volume}
  {99}},\ \bibinfo {pages} {104020} (\bibinfo {year} {2019})},\ \Eprint
  {http://arxiv.org/abs/1812.03420} {arXiv:1812.03420 [gr-qc]} \BibitemShut
  {NoStop}%
\bibitem [{\citenamefont {Takahashi}\ and\ \citenamefont
  {Tenkanen}(2019)}]{Takahashi2019}%
  \BibitemOpen
  \bibfield  {author} {\bibinfo {author} {\bibfnamefont {Tomo}\ \bibnamefont
  {Takahashi}}\ and\ \bibinfo {author} {\bibfnamefont {Tommi}\ \bibnamefont
  {Tenkanen}},\ }\bibfield  {title} {\enquote {\bibinfo {title} {{Towards
  distinguishing variants of non-minimal inflation}},}\ }\href {\doibase
  10.1088/1475-7516/2019/04/035} {\bibfield  {journal} {\bibinfo  {journal}
  {JCAP}\ }\textbf {\bibinfo {volume} {04}},\ \bibinfo {pages} {035} (\bibinfo
  {year} {2019})},\ \Eprint {http://arxiv.org/abs/1812.08492} {arXiv:1812.08492
  [astro-ph.CO]} \BibitemShut {NoStop}%
\bibitem [{\citenamefont {Jinno}\ \emph {et~al.}(2019)\citenamefont {Jinno},
  \citenamefont {Kaneta}, \citenamefont {Oda},\ and\ \citenamefont
  {Park}}]{Jinno2019}%
  \BibitemOpen
  \bibfield  {author} {\bibinfo {author} {\bibfnamefont {Ryusuke}\ \bibnamefont
  {Jinno}}, \bibinfo {author} {\bibfnamefont {Kunio}\ \bibnamefont {Kaneta}},
  \bibinfo {author} {\bibfnamefont {Kin-ya}\ \bibnamefont {Oda}}, \ and\
  \bibinfo {author} {\bibfnamefont {Seong~Chan}\ \bibnamefont {Park}},\
  }\bibfield  {title} {\enquote {\bibinfo {title} {{Hillclimbing inflation in
  metric and Palatini formulations}},}\ }\href {\doibase
  10.1016/j.physletb.2019.03.012} {\bibfield  {journal} {\bibinfo  {journal}
  {Phys. Lett. B}\ }\textbf {\bibinfo {volume} {791}},\ \bibinfo {pages}
  {396--402} (\bibinfo {year} {2019})},\ \Eprint
  {http://arxiv.org/abs/1812.11077} {arXiv:1812.11077 [gr-qc]} \BibitemShut
  {NoStop}%
\bibitem [{\citenamefont {Tenkanen}(2019)}]{Tenkanen2019}%
  \BibitemOpen
  \bibfield  {author} {\bibinfo {author} {\bibfnamefont {Tommi}\ \bibnamefont
  {Tenkanen}},\ }\bibfield  {title} {\enquote {\bibinfo {title} {{Minimal Higgs
  inflation with an $R^2$ term in Palatini gravity}},}\ }\href {\doibase
  10.1103/PhysRevD.99.063528} {\bibfield  {journal} {\bibinfo  {journal} {Phys.
  Rev. D}\ }\textbf {\bibinfo {volume} {99}},\ \bibinfo {pages} {063528}
  (\bibinfo {year} {2019})},\ \Eprint {http://arxiv.org/abs/1901.01794}
  {arXiv:1901.01794 [astro-ph.CO]} \BibitemShut {NoStop}%
\bibitem [{\citenamefont {Edery}\ and\ \citenamefont
  {Nakayama}(2019)}]{Edery2019}%
  \BibitemOpen
  \bibfield  {author} {\bibinfo {author} {\bibfnamefont {Ariel}\ \bibnamefont
  {Edery}}\ and\ \bibinfo {author} {\bibfnamefont {Yu}~\bibnamefont
  {Nakayama}},\ }\bibfield  {title} {\enquote {\bibinfo {title} {{Palatini
  formulation of pure $R^2$ gravity yields Einstein gravity with no massless
  scalar}},}\ }\href {\doibase 10.1103/PhysRevD.99.124018} {\bibfield
  {journal} {\bibinfo  {journal} {Phys. Rev. D}\ }\textbf {\bibinfo {volume}
  {99}},\ \bibinfo {pages} {124018} (\bibinfo {year} {2019})},\ \Eprint
  {http://arxiv.org/abs/1902.07876} {arXiv:1902.07876 [hep-th]} \BibitemShut
  {NoStop}%
\bibitem [{\citenamefont {Jinno}\ \emph {et~al.}(2020)\citenamefont {Jinno},
  \citenamefont {Kubota}, \citenamefont {Oda},\ and\ \citenamefont
  {Park}}]{Jinno2020}%
  \BibitemOpen
  \bibfield  {author} {\bibinfo {author} {\bibfnamefont {Ryusuke}\ \bibnamefont
  {Jinno}}, \bibinfo {author} {\bibfnamefont {Mio}\ \bibnamefont {Kubota}},
  \bibinfo {author} {\bibfnamefont {Kin-ya}\ \bibnamefont {Oda}}, \ and\
  \bibinfo {author} {\bibfnamefont {Seong~Chan}\ \bibnamefont {Park}},\
  }\bibfield  {title} {\enquote {\bibinfo {title} {{Higgs inflation in metric
  and Palatini formalisms: Required suppression of higher dimensional
  operators}},}\ }\href {\doibase 10.1088/1475-7516/2020/03/063} {\bibfield
  {journal} {\bibinfo  {journal} {JCAP}\ }\textbf {\bibinfo {volume} {03}},\
  \bibinfo {pages} {063} (\bibinfo {year} {2020})},\ \Eprint
  {http://arxiv.org/abs/1904.05699} {arXiv:1904.05699 [hep-ph]} \BibitemShut
  {NoStop}%
\bibitem [{\citenamefont {Aoki}\ and\ \citenamefont
  {Shimada}(2019)}]{Aoki2019}%
  \BibitemOpen
  \bibfield  {author} {\bibinfo {author} {\bibfnamefont {Katsuki}\ \bibnamefont
  {Aoki}}\ and\ \bibinfo {author} {\bibfnamefont {Keigo}\ \bibnamefont
  {Shimada}},\ }\bibfield  {title} {\enquote {\bibinfo {title}
  {{Scalar-metric-affine theories: Can we get ghost-free theories from
  symmetry?}}}\ }\href {\doibase 10.1103/PhysRevD.100.044037} {\bibfield
  {journal} {\bibinfo  {journal} {Phys. Rev. D}\ }\textbf {\bibinfo {volume}
  {100}},\ \bibinfo {pages} {044037} (\bibinfo {year} {2019})},\ \Eprint
  {http://arxiv.org/abs/1904.10175} {arXiv:1904.10175 [hep-th]} \BibitemShut
  {NoStop}%
\bibitem [{\citenamefont {Giovannini}(2019)}]{Giovannini2019}%
  \BibitemOpen
  \bibfield  {author} {\bibinfo {author} {\bibfnamefont {Massimo}\ \bibnamefont
  {Giovannini}},\ }\bibfield  {title} {\enquote {\bibinfo {title}
  {{Post-inflationary phases stiffer than radiation and Palatini
  formulation}},}\ }\href {\doibase 10.1088/1361-6382/ab52a8} {\bibfield
  {journal} {\bibinfo  {journal} {Class. Quant. Grav.}\ }\textbf {\bibinfo
  {volume} {36}},\ \bibinfo {pages} {235017} (\bibinfo {year} {2019})},\
  \Eprint {http://arxiv.org/abs/1905.06182} {arXiv:1905.06182 [gr-qc]}
  \BibitemShut {NoStop}%
\bibitem [{\citenamefont {Tenkanen}\ and\ \citenamefont
  {Visinelli}(2019)}]{Tenkanen2019a}%
  \BibitemOpen
  \bibfield  {author} {\bibinfo {author} {\bibfnamefont {Tommi}\ \bibnamefont
  {Tenkanen}}\ and\ \bibinfo {author} {\bibfnamefont {Luca}\ \bibnamefont
  {Visinelli}},\ }\bibfield  {title} {\enquote {\bibinfo {title} {{Axion dark
  matter from Higgs inflation with an intermediate $H_*$}},}\ }\href {\doibase
  10.1088/1475-7516/2019/08/033} {\bibfield  {journal} {\bibinfo  {journal}
  {JCAP}\ }\textbf {\bibinfo {volume} {08}},\ \bibinfo {pages} {033} (\bibinfo
  {year} {2019})},\ \Eprint {http://arxiv.org/abs/1906.11837} {arXiv:1906.11837
  [astro-ph.CO]} \BibitemShut {NoStop}%
\bibitem [{\citenamefont {Bostan}(2019{\natexlab{a}})}]{Bostan2019a}%
  \BibitemOpen
  \bibfield  {author} {\bibinfo {author} {\bibfnamefont {Nilay}\ \bibnamefont
  {Bostan}},\ }\bibfield  {title} {\enquote {\bibinfo {title} {{Non-minimally
  coupled quartic inflation with Coleman-Weinberg one-loop corrections in the
  Palatini formulation}},}\ }\href@noop {} {\  (\bibinfo {year}
  {2019}{\natexlab{a}})},\ \Eprint {http://arxiv.org/abs/1907.13235}
  {arXiv:1907.13235 [gr-qc]} \BibitemShut {NoStop}%
\bibitem [{\citenamefont {Bostan}(2019{\natexlab{b}})}]{Bostan2019}%
  \BibitemOpen
  \bibfield  {author} {\bibinfo {author} {\bibfnamefont {Nilay}\ \bibnamefont
  {Bostan}},\ }\bibfield  {title} {\enquote {\bibinfo {title} {{Quadratic,
  Higgs and hilltop potentials in the Palatini gravity}},}\ }\href@noop {} {\
  (\bibinfo {year} {2019}{\natexlab{b}})},\ \Eprint
  {http://arxiv.org/abs/1908.09674} {arXiv:1908.09674 [astro-ph.CO]}
  \BibitemShut {NoStop}%
\bibitem [{\citenamefont {Tenkanen}(2020{\natexlab{a}})}]{Tenkanen2020}%
  \BibitemOpen
  \bibfield  {author} {\bibinfo {author} {\bibfnamefont {Tommi}\ \bibnamefont
  {Tenkanen}},\ }\bibfield  {title} {\enquote {\bibinfo {title}
  {{Trans-Planckian censorship, inflation, and dark matter}},}\ }\href
  {\doibase 10.1103/PhysRevD.101.063517} {\bibfield  {journal} {\bibinfo
  {journal} {Phys. Rev. D}\ }\textbf {\bibinfo {volume} {101}},\ \bibinfo
  {pages} {063517} (\bibinfo {year} {2020}{\natexlab{a}})},\ \Eprint
  {http://arxiv.org/abs/1910.00521} {arXiv:1910.00521 [astro-ph.CO]}
  \BibitemShut {NoStop}%
\bibitem [{\citenamefont {Gialamas}\ and\ \citenamefont
  {Lahanas}(2020)}]{Gialamas2020}%
  \BibitemOpen
  \bibfield  {author} {\bibinfo {author} {\bibfnamefont {Ioannis~D.}\
  \bibnamefont {Gialamas}}\ and\ \bibinfo {author} {\bibfnamefont {A.B.}\
  \bibnamefont {Lahanas}},\ }\bibfield  {title} {\enquote {\bibinfo {title}
  {{Reheating in $R^2$ Palatini inflationary models}},}\ }\href {\doibase
  10.1103/PhysRevD.101.084007} {\bibfield  {journal} {\bibinfo  {journal}
  {Phys. Rev. D}\ }\textbf {\bibinfo {volume} {101}},\ \bibinfo {pages}
  {084007} (\bibinfo {year} {2020})},\ \Eprint
  {http://arxiv.org/abs/1911.11513} {arXiv:1911.11513 [gr-qc]} \BibitemShut
  {NoStop}%
\bibitem [{\citenamefont {Racioppi}(2019)}]{Racioppi2019}%
  \BibitemOpen
  \bibfield  {author} {\bibinfo {author} {\bibfnamefont {Antonio}\ \bibnamefont
  {Racioppi}},\ }\bibfield  {title} {\enquote {\bibinfo {title} {{Non-Minimal
  (Self-)Running Inflation: Metric vs. Palatini Formulation}},}\ }\href@noop {}
  {\  (\bibinfo {year} {2019})},\ \Eprint {http://arxiv.org/abs/1912.10038}
  {arXiv:1912.10038 [hep-ph]} \BibitemShut {NoStop}%
\bibitem [{\citenamefont {Antoniadis}\ \emph {et~al.}(2018)\citenamefont
  {Antoniadis}, \citenamefont {Karam}, \citenamefont {Lykkas},\ and\
  \citenamefont {Tamvakis}}]{Antoniadis2018}%
  \BibitemOpen
  \bibfield  {author} {\bibinfo {author} {\bibfnamefont {I.}~\bibnamefont
  {Antoniadis}}, \bibinfo {author} {\bibfnamefont {A.}~\bibnamefont {Karam}},
  \bibinfo {author} {\bibfnamefont {A.}~\bibnamefont {Lykkas}}, \ and\ \bibinfo
  {author} {\bibfnamefont {K.}~\bibnamefont {Tamvakis}},\ }\bibfield  {title}
  {\enquote {\bibinfo {title} {{Palatini inflation in models with an $R^2$
  term}},}\ }\href {\doibase 10.1088/1475-7516/2018/11/028} {\bibfield
  {journal} {\bibinfo  {journal} {JCAP}\ }\textbf {\bibinfo {volume} {1811}},\
  \bibinfo {pages} {028} (\bibinfo {year} {2018})},\ \Eprint
  {http://arxiv.org/abs/1810.10418} {arXiv:1810.10418 [gr-qc]} \BibitemShut
  {NoStop}%
\bibitem [{\citenamefont {Antoniadis}\ \emph
  {et~al.}(2019{\natexlab{a}})\citenamefont {Antoniadis}, \citenamefont
  {Karam}, \citenamefont {Lykkas}, \citenamefont {Pappas},\ and\ \citenamefont
  {Tamvakis}}]{Antoniadis2019}%
  \BibitemOpen
  \bibfield  {author} {\bibinfo {author} {\bibfnamefont {I.}~\bibnamefont
  {Antoniadis}}, \bibinfo {author} {\bibfnamefont {A.}~\bibnamefont {Karam}},
  \bibinfo {author} {\bibfnamefont {A.}~\bibnamefont {Lykkas}}, \bibinfo
  {author} {\bibfnamefont {T.}~\bibnamefont {Pappas}}, \ and\ \bibinfo {author}
  {\bibfnamefont {K.}~\bibnamefont {Tamvakis}},\ }\bibfield  {title} {\enquote
  {\bibinfo {title} {{Rescuing Quartic and Natural Inflation in the Palatini
  Formalism}},}\ }\href {\doibase 10.1088/1475-7516/2019/03/005} {\bibfield
  {journal} {\bibinfo  {journal} {JCAP}\ }\textbf {\bibinfo {volume} {1903}},\
  \bibinfo {pages} {005} (\bibinfo {year} {2019}{\natexlab{a}})},\ \Eprint
  {http://arxiv.org/abs/1812.00847} {arXiv:1812.00847 [gr-qc]} \BibitemShut
  {NoStop}%
\bibitem [{\citenamefont {Antoniadis}\ \emph
  {et~al.}(2019{\natexlab{b}})\citenamefont {Antoniadis}, \citenamefont
  {Karam}, \citenamefont {Lykkas}, \citenamefont {Pappas},\ and\ \citenamefont
  {Tamvakis}}]{Antoniadis2019a}%
  \BibitemOpen
  \bibfield  {author} {\bibinfo {author} {\bibfnamefont {Ignatios}\
  \bibnamefont {Antoniadis}}, \bibinfo {author} {\bibfnamefont {Alexandros}\
  \bibnamefont {Karam}}, \bibinfo {author} {\bibfnamefont {Angelos}\
  \bibnamefont {Lykkas}}, \bibinfo {author} {\bibfnamefont {Thomas}\
  \bibnamefont {Pappas}}, \ and\ \bibinfo {author} {\bibfnamefont {Kyriakos}\
  \bibnamefont {Tamvakis}},\ }\bibfield  {title} {\enquote {\bibinfo {title}
  {{Single-field inflation in models with an $R^2$ term}},}\ }in\ \href@noop {}
  {\emph {\bibinfo {booktitle} {{19th Hellenic School and Workshops on
  Elementary Particle Physics and Gravity}}}}\ (\bibinfo {year} {2019})\
  \Eprint {http://arxiv.org/abs/1912.12757} {arXiv:1912.12757 [gr-qc]}
  \BibitemShut {NoStop}%
\bibitem [{\citenamefont {Tenkanen}(2020{\natexlab{b}})}]{Tenkanen2020a}%
  \BibitemOpen
  \bibfield  {author} {\bibinfo {author} {\bibfnamefont {Tommi}\ \bibnamefont
  {Tenkanen}},\ }\bibfield  {title} {\enquote {\bibinfo {title} {{Tracing the
  high energy theory of gravity: an introduction to Palatini inflation}},}\
  }\href {\doibase 10.1007/s10714-020-02682-2} {\bibfield  {journal} {\bibinfo
  {journal} {Gen. Rel. Grav.}\ }\textbf {\bibinfo {volume} {52}},\ \bibinfo
  {pages} {33} (\bibinfo {year} {2020}{\natexlab{b}})},\ \Eprint
  {http://arxiv.org/abs/2001.10135} {arXiv:2001.10135 [astro-ph.CO]}
  \BibitemShut {NoStop}%
\bibitem [{\citenamefont {Tenkanen}\ and\ \citenamefont
  {Tomberg}(2020)}]{Tenkanen2020b}%
  \BibitemOpen
  \bibfield  {author} {\bibinfo {author} {\bibfnamefont {Tommi}\ \bibnamefont
  {Tenkanen}}\ and\ \bibinfo {author} {\bibfnamefont {Eemeli}\ \bibnamefont
  {Tomberg}},\ }\bibfield  {title} {\enquote {\bibinfo {title} {{Initial
  conditions for plateau inflation: a case study}},}\ }\href {\doibase
  10.1088/1475-7516/2020/04/050} {\  (\bibinfo {year} {2020}),\
  10.1088/1475-7516/2020/04/050},\ \Eprint {http://arxiv.org/abs/2002.02420}
  {arXiv:2002.02420 [astro-ph.CO]} \BibitemShut {NoStop}%
\bibitem [{\citenamefont {Lloyd-Stubbs}\ and\ \citenamefont
  {McDonald}(2020)}]{LloydStubbs2020}%
  \BibitemOpen
  \bibfield  {author} {\bibinfo {author} {\bibfnamefont {Amy}\ \bibnamefont
  {Lloyd-Stubbs}}\ and\ \bibinfo {author} {\bibfnamefont {John}\ \bibnamefont
  {McDonald}},\ }\bibfield  {title} {\enquote {\bibinfo {title} {{Sub-Planckian
  $\phi^{2}$ Inflation in the Palatini Formulation of Gravity with an $R^2$
  term}},}\ }\href@noop {} {\  (\bibinfo {year} {2020})},\ \Eprint
  {http://arxiv.org/abs/2002.08324} {arXiv:2002.08324 [hep-ph]} \BibitemShut
  {NoStop}%
\bibitem [{\citenamefont {Antoniadis}\ \emph {et~al.}(2020)\citenamefont
  {Antoniadis}, \citenamefont {Lykkas},\ and\ \citenamefont
  {Tamvakis}}]{Antoniadis2020}%
  \BibitemOpen
  \bibfield  {author} {\bibinfo {author} {\bibfnamefont {Ignatios}\
  \bibnamefont {Antoniadis}}, \bibinfo {author} {\bibfnamefont {Angelos}\
  \bibnamefont {Lykkas}}, \ and\ \bibinfo {author} {\bibfnamefont {Kyriakos}\
  \bibnamefont {Tamvakis}},\ }\bibfield  {title} {\enquote {\bibinfo {title}
  {{Constant-roll in the Palatini-$R^2$ models}},}\ }\href {\doibase
  10.1088/1475-7516/2020/04/033} {\bibfield  {journal} {\bibinfo  {journal}
  {JCAP}\ }\textbf {\bibinfo {volume} {04}},\ \bibinfo {pages} {033} (\bibinfo
  {year} {2020})},\ \Eprint {http://arxiv.org/abs/2002.12681} {arXiv:2002.12681
  [gr-qc]} \BibitemShut {NoStop}%
\bibitem [{\citenamefont {Ghilencea}(2020)}]{Ghilencea2020}%
  \BibitemOpen
  \bibfield  {author} {\bibinfo {author} {\bibfnamefont {D.M.}\ \bibnamefont
  {Ghilencea}},\ }\bibfield  {title} {\enquote {\bibinfo {title} {{Palatini
  quadratic gravity: spontaneous breaking of gauged scale symmetry and
  inflation}},}\ }\href@noop {} {\  (\bibinfo {year} {2020})},\ \Eprint
  {http://arxiv.org/abs/2003.08516} {arXiv:2003.08516 [hep-th]} \BibitemShut
  {NoStop}%
\bibitem [{\citenamefont {Matsumura}\ \emph {et~al.}(2016)\citenamefont
  {Matsumura} \emph {et~al.}}]{Matsumura2016}%
  \BibitemOpen
  \bibfield  {author} {\bibinfo {author} {\bibfnamefont {T.}~\bibnamefont
  {Matsumura}} \emph {et~al.},\ }\bibfield  {title} {\enquote {\bibinfo {title}
  {Litebird: Mission overview and focal plane layout},}\ }\href {\doibase
  10.1007/s10909-016-1542-8} {\bibfield  {journal} {\bibinfo  {journal}
  {Journal of Low Temperature Physics}\ }\textbf {\bibinfo {volume} {184}},\
  \bibinfo {pages} {824--831} (\bibinfo {year} {2016})}\BibitemShut {NoStop}%
\bibitem [{\citenamefont {Kogut}\ \emph {et~al.}(2011)\citenamefont {Kogut},
  \citenamefont {Fixsen}, \citenamefont {Chuss}, \citenamefont {Dotson},
  \citenamefont {Dwek}, \citenamefont {Halpern}, \citenamefont {Hinshaw},
  \citenamefont {Meyer}, \citenamefont {Moseley}, \citenamefont {Seiffert},
  \citenamefont {Spergel},\ and\ \citenamefont {Wollack}}]{Kogut_2011}%
  \BibitemOpen
  \bibfield  {author} {\bibinfo {author} {\bibfnamefont {A}~\bibnamefont
  {Kogut}}, \bibinfo {author} {\bibfnamefont {D.J}\ \bibnamefont {Fixsen}},
  \bibinfo {author} {\bibfnamefont {D.T}\ \bibnamefont {Chuss}}, \bibinfo
  {author} {\bibfnamefont {J}~\bibnamefont {Dotson}}, \bibinfo {author}
  {\bibfnamefont {E}~\bibnamefont {Dwek}}, \bibinfo {author} {\bibfnamefont
  {M}~\bibnamefont {Halpern}}, \bibinfo {author} {\bibfnamefont {G.F}\
  \bibnamefont {Hinshaw}}, \bibinfo {author} {\bibfnamefont {S.M}\ \bibnamefont
  {Meyer}}, \bibinfo {author} {\bibfnamefont {S.H}\ \bibnamefont {Moseley}},
  \bibinfo {author} {\bibfnamefont {M.D}\ \bibnamefont {Seiffert}}, \bibinfo
  {author} {\bibfnamefont {D.N}\ \bibnamefont {Spergel}}, \ and\ \bibinfo
  {author} {\bibfnamefont {E.J}\ \bibnamefont {Wollack}},\ }\bibfield  {title}
  {\enquote {\bibinfo {title} {The primordial inflation explorer ({PIXIE}): a
  nulling polarimeter for cosmic microwave background observations},}\ }\href
  {\doibase 10.1088/1475-7516/2011/07/025} {\bibfield  {journal} {\bibinfo
  {journal} {Journal of Cosmology and Astroparticle Physics}\ }\textbf
  {\bibinfo {volume} {2011}},\ \bibinfo {pages} {025--025} (\bibinfo {year}
  {2011})}\BibitemShut {NoStop}%
\bibitem [{\citenamefont {Sutin}\ \emph {et~al.}(2018)\citenamefont {Sutin}
  \emph {et~al.}}]{Sutin:2018onu}%
  \BibitemOpen
  \bibfield  {author} {\bibinfo {author} {\bibfnamefont {Brian~M.}\
  \bibnamefont {Sutin}} \emph {et~al.},\ }\bibfield  {title} {\enquote
  {\bibinfo {title} {{PICO - the probe of inflation and cosmic origins}},}\
  }\bibfield  {booktitle} {\emph {\bibinfo {booktitle} {{Proceedings, SPIE
  Astronomical Telescopes + Instrumentation 2018: Modeling, Systems
  Engineering, and Project Management for Astronomy VIII: Austin, USA, June
  10-15, 2018}}},\ }\href {\doibase 10.1117/12.2311326} {\bibfield  {journal}
  {\bibinfo  {journal} {Proc. SPIE Int. Soc. Opt. Eng.}\ }\textbf {\bibinfo
  {volume} {10698}},\ \bibinfo {pages} {106984F} (\bibinfo {year} {2018})},\
  \Eprint {http://arxiv.org/abs/1808.01368} {arXiv:1808.01368 [astro-ph.IM]}
  \BibitemShut {NoStop}%
\bibitem [{\citenamefont {Flanagan}(2004)}]{Flanagan:2004bz}%
  \BibitemOpen
  \bibfield  {author} {\bibinfo {author} {\bibfnamefont {Eanna~E.}\
  \bibnamefont {Flanagan}},\ }\bibfield  {title} {\enquote {\bibinfo {title}
  {{The Conformal frame freedom in theories of gravitation}},}\ }\href
  {\doibase 10.1088/0264-9381/21/15/N02} {\bibfield  {journal} {\bibinfo
  {journal} {Class. Quant. Grav.}\ }\textbf {\bibinfo {volume} {21}},\ \bibinfo
  {pages} {3817} (\bibinfo {year} {2004})},\ \Eprint
  {http://arxiv.org/abs/gr-qc/0403063} {arXiv:gr-qc/0403063 [gr-qc]}
  \BibitemShut {NoStop}%
\bibitem [{\citenamefont {Rubio}\ and\ \citenamefont
  {Tomberg}(2019)}]{Rubio:2019ypq}%
  \BibitemOpen
  \bibfield  {author} {\bibinfo {author} {\bibfnamefont {Javier}\ \bibnamefont
  {Rubio}}\ and\ \bibinfo {author} {\bibfnamefont {Eemeli~S.}\ \bibnamefont
  {Tomberg}},\ }\bibfield  {title} {\enquote {\bibinfo {title} {{Preheating in
  Palatini Higgs inflation}},}\ }\href {\doibase 10.1088/1475-7516/2019/04/021}
  {\bibfield  {journal} {\bibinfo  {journal} {JCAP}\ }\textbf {\bibinfo
  {volume} {04}},\ \bibinfo {pages} {021} (\bibinfo {year} {2019})},\ \Eprint
  {http://arxiv.org/abs/1902.10148} {arXiv:1902.10148 [hep-ph]} \BibitemShut
  {NoStop}%
\bibitem [{\citenamefont {Karamitsos}(2019)}]{Karamitsos:2019vor}%
  \BibitemOpen
  \bibfield  {author} {\bibinfo {author} {\bibfnamefont {Sotirios}\
  \bibnamefont {Karamitsos}},\ }\bibfield  {title} {\enquote {\bibinfo {title}
  {{Beyond the Poles in Attractor Models of Inflation}},}\ }\href {\doibase
  10.1088/1475-7516/2019/09/022} {\bibfield  {journal} {\bibinfo  {journal}
  {JCAP}\ }\textbf {\bibinfo {volume} {09}},\ \bibinfo {pages} {022} (\bibinfo
  {year} {2019})},\ \Eprint {http://arxiv.org/abs/1903.03707} {arXiv:1903.03707
  [hep-th]} \BibitemShut {NoStop}%
\bibitem [{\citenamefont {Jarv}\ \emph {et~al.}(2010)\citenamefont {Jarv},
  \citenamefont {Kuusk},\ and\ \citenamefont {Saal}}]{Jarv:2010zc}%
  \BibitemOpen
  \bibfield  {author} {\bibinfo {author} {\bibfnamefont {Laur}\ \bibnamefont
  {Jarv}}, \bibinfo {author} {\bibfnamefont {Piret}\ \bibnamefont {Kuusk}}, \
  and\ \bibinfo {author} {\bibfnamefont {Margus}\ \bibnamefont {Saal}},\
  }\bibfield  {title} {\enquote {\bibinfo {title} {{Potential dominated
  scalar-tensor cosmologies in the general relativity limit: phase space
  view}},}\ }\href {\doibase 10.1103/PhysRevD.81.104007} {\bibfield  {journal}
  {\bibinfo  {journal} {Phys. Rev. D}\ }\textbf {\bibinfo {volume} {81}},\
  \bibinfo {pages} {104007} (\bibinfo {year} {2010})},\ \Eprint
  {http://arxiv.org/abs/1003.1686} {arXiv:1003.1686 [gr-qc]} \BibitemShut
  {NoStop}%
\bibitem [{\citenamefont {Briscese}\ \emph {et~al.}(2007)\citenamefont
  {Briscese}, \citenamefont {Elizalde}, \citenamefont {Nojiri},\ and\
  \citenamefont {Odintsov}}]{Briscese:2006xu}%
  \BibitemOpen
  \bibfield  {author} {\bibinfo {author} {\bibfnamefont {F.}~\bibnamefont
  {Briscese}}, \bibinfo {author} {\bibfnamefont {E.}~\bibnamefont {Elizalde}},
  \bibinfo {author} {\bibfnamefont {S.}~\bibnamefont {Nojiri}}, \ and\ \bibinfo
  {author} {\bibfnamefont {S.D.}\ \bibnamefont {Odintsov}},\ }\bibfield
  {title} {\enquote {\bibinfo {title} {{Phantom scalar dark energy as modified
  gravity: Understanding the origin of the Big Rip singularity}},}\ }\href
  {\doibase 10.1016/j.physletb.2007.01.013} {\bibfield  {journal} {\bibinfo
  {journal} {Phys. Lett. B}\ }\textbf {\bibinfo {volume} {646}},\ \bibinfo
  {pages} {105--111} (\bibinfo {year} {2007})},\ \Eprint
  {http://arxiv.org/abs/hep-th/0612220} {arXiv:hep-th/0612220} \BibitemShut
  {NoStop}%
\bibitem [{\citenamefont {Bahamonde}\ \emph {et~al.}(2016)\citenamefont
  {Bahamonde}, \citenamefont {Odintsov}, \citenamefont {Oikonomou},\ and\
  \citenamefont {Wright}}]{Bahamonde:2016wmz}%
  \BibitemOpen
  \bibfield  {author} {\bibinfo {author} {\bibfnamefont {Sebastian}\
  \bibnamefont {Bahamonde}}, \bibinfo {author} {\bibfnamefont {S.D.}\
  \bibnamefont {Odintsov}}, \bibinfo {author} {\bibfnamefont {V.K.}\
  \bibnamefont {Oikonomou}}, \ and\ \bibinfo {author} {\bibfnamefont {Matthew}\
  \bibnamefont {Wright}},\ }\bibfield  {title} {\enquote {\bibinfo {title}
  {{Correspondence of $F(R)$ Gravity Singularities in Jordan and Einstein
  Frames}},}\ }\href {\doibase 10.1016/j.aop.2016.06.020} {\bibfield  {journal}
  {\bibinfo  {journal} {Annals Phys.}\ }\textbf {\bibinfo {volume} {373}},\
  \bibinfo {pages} {96--114} (\bibinfo {year} {2016})},\ \Eprint
  {http://arxiv.org/abs/1603.05113} {arXiv:1603.05113 [gr-qc]} \BibitemShut
  {NoStop}%
\bibitem [{\citenamefont {Ferrara}\ \emph {et~al.}(2013)\citenamefont
  {Ferrara}, \citenamefont {Kallosh}, \citenamefont {Linde},\ and\
  \citenamefont {Porrati}}]{Ferrara2013b}%
  \BibitemOpen
  \bibfield  {author} {\bibinfo {author} {\bibfnamefont {Sergio}\ \bibnamefont
  {Ferrara}}, \bibinfo {author} {\bibfnamefont {Renata}\ \bibnamefont
  {Kallosh}}, \bibinfo {author} {\bibfnamefont {Andrei}\ \bibnamefont {Linde}},
  \ and\ \bibinfo {author} {\bibfnamefont {Massimo}\ \bibnamefont {Porrati}},\
  }\bibfield  {title} {\enquote {\bibinfo {title} {{Minimal Supergravity Models
  of Inflation}},}\ }\href {\doibase 10.1103/PhysRevD.88.085038} {\bibfield
  {journal} {\bibinfo  {journal} {Phys. Rev.}\ }\textbf {\bibinfo {volume}
  {D88}},\ \bibinfo {pages} {085038} (\bibinfo {year} {2013})},\ \Eprint
  {http://arxiv.org/abs/1307.7696} {arXiv:1307.7696 [hep-th]} \BibitemShut
  {NoStop}%
\bibitem [{\citenamefont {Kallosh}\ \emph {et~al.}(2013)\citenamefont
  {Kallosh}, \citenamefont {Linde},\ and\ \citenamefont
  {Roest}}]{Kallosh2013a}%
  \BibitemOpen
  \bibfield  {author} {\bibinfo {author} {\bibfnamefont {Renata}\ \bibnamefont
  {Kallosh}}, \bibinfo {author} {\bibfnamefont {Andrei}\ \bibnamefont {Linde}},
  \ and\ \bibinfo {author} {\bibfnamefont {Diederik}\ \bibnamefont {Roest}},\
  }\bibfield  {title} {\enquote {\bibinfo {title} {{Superconformal Inflationary
  $\alpha$-Attractors}},}\ }\href {\doibase 10.1007/JHEP11(2013)198} {\bibfield
   {journal} {\bibinfo  {journal} {JHEP}\ }\textbf {\bibinfo {volume} {11}},\
  \bibinfo {pages} {198} (\bibinfo {year} {2013})},\ \Eprint
  {http://arxiv.org/abs/1311.0472} {arXiv:1311.0472 [hep-th]} \BibitemShut
  {NoStop}%
\bibitem [{\citenamefont {Galante}\ \emph {et~al.}(2015)\citenamefont
  {Galante}, \citenamefont {Kallosh}, \citenamefont {Linde},\ and\
  \citenamefont {Roest}}]{Galante2015}%
  \BibitemOpen
  \bibfield  {author} {\bibinfo {author} {\bibfnamefont {Mario}\ \bibnamefont
  {Galante}}, \bibinfo {author} {\bibfnamefont {Renata}\ \bibnamefont
  {Kallosh}}, \bibinfo {author} {\bibfnamefont {Andrei}\ \bibnamefont {Linde}},
  \ and\ \bibinfo {author} {\bibfnamefont {Diederik}\ \bibnamefont {Roest}},\
  }\bibfield  {title} {\enquote {\bibinfo {title} {{Unity of Cosmological
  Inflation Attractors}},}\ }\href {\doibase 10.1103/PhysRevLett.114.141302}
  {\bibfield  {journal} {\bibinfo  {journal} {Phys. Rev. Lett.}\ }\textbf
  {\bibinfo {volume} {114}},\ \bibinfo {pages} {141302} (\bibinfo {year}
  {2015})},\ \Eprint {http://arxiv.org/abs/1412.3797} {arXiv:1412.3797
  [hep-th]} \BibitemShut {NoStop}%
\end{thebibliography}%

\end{document}